\documentclass[12pt]{JHEP3}
\usepackage{amsfonts}
\usepackage{amscd}
\usepackage{amssymb}
\usepackage{amsmath}
\usepackage{cite}
\usepackage{latexsym}

\renewcommand{\j}{j}

\renewcommand{\Re}{{\rm Re}}
\renewcommand{\tilde}{\widetilde}
\newcommand{\ov}{\overline}
\newcommand{\be}{\begin{equation}}
\newcommand{\ee}{\end{equation}}
\newcommand{\ba}{\begin{eqnarray}}
\newcommand{\ea}{\end{eqnarray}}
\newcommand{\del}{\partial}
\newcommand{\delbar}{\overline{\del}}

\newcommand{\bC}{{\mathbb{C}}}

\newcommand{\bR}{{\mathbb{R}}}
\newcommand{\bP}{{\mathbb{P}}}
\newcommand{\bN}{{\mathbb{PN}}}

\newcommand{\bT}{{\mathbb{PT}}}

\newcommand{\bZ}{{\mathbb{Z}}}
\newcommand{\cA}{{\mathcal{A}}}
\newcommand{\cB}{{\mathcal{B}}}

\newcommand{\cD}{{\mathcal{D}}}

\newcommand{\cH}{{\mathcal{H}}}

\newcommand{\cJ}{{\mathcal{J}}}
\newcommand{\cK}{{\mathcal{K}}}
\newcommand{\cL}{{\mathcal{L}}}
\newcommand{\cM}{{\mathcal{M}}}
\newcommand{\cN}{{\mathcal{N}}}
\newcommand{\cO}{{\mathcal{O}}}
\newcommand{\cS}{{\mathcal{S}}}
\newcommand{\cV}{{\mathcal{V}}}
\newcommand{\cW}{{\mathcal{W}}}

\newcommand{\rd}{{\rm d}}

\newcommand{\e}{{\rm e}}
\newcommand{\im}{{\rm i}}

\title{Heterotic Twistor-String Theory}
\author{Lionel Mason and David Skinner\\
The Mathematical Institute, University of Oxford \\
24-29 St. Giles, Oxford OX1 3LP, United Kingdom \\
{\rm \{lmason, skinnerd\}@maths.ox.ac.uk}}

\abstract{We reformulate twistor-string theory as a heterotic string
based on a twisted (0,2) model. The path integral localizes on
holomorphic maps, while the (0,2) moduli naturally correspond to the
states of $\cN=4$ super Yang-Mills and conformal supergravity under
the Penrose transform. We show how the standard twistor-string
formulae of scattering amplitudes as integrals over the space of
curves in supertwistor space may be obtained from our model.
The corresponding string field theory gives rise to a twistor
action for $\cN=4$ conformal supergravity coupled to super
Yang-Mills. The model helps to explain how the twistor-strings of
Witten and Berkovits are related and clarifies various aspects of each of these models.}

\begin{document}

\section{Introduction}

The twistor-string theories of Witten~\cite{Witten:2003nn} and
Berkovits~\cite{Berkovits:2004hg} combine topological string theory
with the Penrose transform~\cite{Penrose:1972ia} to describe field
theories in four dimensional spacetime. The models appear to be
equivalent to each other and to $\cN=4$ super~Yang-Mills theory
coupled to a non-minimal conformal
supergravity~\cite{Berkovits:2004jj}.  The mechanism is completely
different from the usual string paradigm: spacetime is not
introduced {\it ab initio} as a target, but emerges as the space of
degree 1 worldsheet instantons in the twistor space target. It
therefore provides a new way for both string theory and twistor
theory to make contact with spacetime physics.  As far as string
theory is concerned, it does so without the extra spacetime
dimensions and further infinite towers of massive modes of
conventional string theory.  As far as twistor theory is concerned,
it resolves (albeit perturbatively) the most serious outstanding
questions in the twistor programme. Firstly, it provides a solution
to the `googly problem' of encoding both the selfdual and
anti-selfdual parts of Yang-Mills and gravitational fields on
twistor space in such a way that interactions can be naturally
incorporated. Classical twistor constructions have previously only
been able to cope with anti-selfdual interactions.  Secondly,
twistor-string theory also provides a natural way to incorporate
quantum field theory into twistor theory.  Moreover the associated
twistor-string field theory is closely related to the twistor
actions constructed in~\cite{Mason:2005zm,Boels:2006ir,Boels:2007qn}. These actions provide generating
principles for all the amplitudes in the theories. Insight from the
twistor-string has also led to a number of powerful new approaches
to calculating scattering amplitudes in perturbative gauge theory,
both directly in string
theory~\cite{Roiban:2004ka,Roiban:2004yf,Gukov:2004ei}, and
indirectly through spacetime unitarity methods inspired by the
twistor-string~\cite{Cachazo:2004kj,Brandhuber:2004yw,Bena:2004xu,Britto:2004nc,Britto:2004ap,
Bern:2005iz,Bern:2005cq}.

There remain a number of difficulties in making sense of
twistor-string theory, and in exploiting it as a calculational tool.
In particular, the presence of conformal supergravity limits ones
ability to use twistor-string theory to calculate pure Yang-Mills
amplitudes to tree level, since supergravity modes will propagate in
any loops~\cite{Witten:2003nn,Dolan:2007vv}. Conformal supergravity is
thought neither to be unitary, nor to possess a stable
vacuum~\cite{Fradkin:1985am} and so is widely viewed as an unwelcome
feature of twistor-string theory. However, because conformal
supergravity contains Poincar{\'e} supergravity as a subsector, one
might more optimistically view it as an opportunity. Indeed,
twistor-string theories with the spectrum of Poincar{\'e} supergravity
have been constructed in~\cite{AbouZeid:2006wu}, although these theories
remain tentative as it has not yet been determined whether they lead
to the correct interactions.  If they do, and are consistent, they will
provide a new approach to quantum gravity.  Furthermore, for
applications to loop calculations in gauge theories, one might then
decouple gravity in the limit that the Planck mass becomes infinite
while the gauge coupling stays finite.

This paper will not attempt to make further progress on these issues,
but will provide a new model for twistor-string theory that goes some
way towards resolving other puzzles arising from the original models.
Witten's original twistor-string~\cite{Witten:2003nn} is based on a
topological string theory, the B-model, of maps from a Riemann surface
into the twistor superspace $\bP^{3|4}$, the projectivization of
$\bC^{4|4}$ with four bosonic coordinates and four fermionic. While
one can always construct a topological string theory on a standard
(bosonic) Calabi-Yau threefold~\cite{Witten:1992fb,Bershadsky:1993cx},
it is not obvious that the same construction works on a supermanifold
such as $\bP^{3|4}$ even if it is formally Calabi-Yau. Proceeding
heuristically, Witten showed that the open string sector would
successfully provide the anti-selfdual\footnote{Our conventions are
  those of Penrose \& Rindler~\cite{Penrose:1986ca}, whereby an
  on-shell massless field of helicity $h$ is represented on twistor
  space $\bT'$ by an element of $H^1(\bT',\cO(-2h-2))$; these
  conventions differ from those of Witten~\cite{Witten:2003nn}.}
interactions of $\cN=4$ super Yang-Mills. However, to include selfdual
interactions requires the introduction of D1-branes wrapping
holomorphic curves in projective supertwistor space. The full
Yang-Mills perturbation theory then arises from strings stretched
between these D1-branes and a stack of (almost) space-filling
D5-branes, together with the holomorphic Chern-Simons theory of the
D5-D5 strings. However, one would also
expect to find open D1-D1 strings and the role of these in spacetime
was left unclear. Gravitational
modes decouple from the open B-model at the perturbative level, so
conformal supergravity arises through the dynamics of the D1-branes in
a manner that was not made entirely transparent. These D-branes are
non-perturbative features of the B-model and thus to fully understand
the presence of conformal supergravity in Witten's model (perhaps so
as to explore related theories with Einstein gravity), one would
appear to have to understand the full non-perturbative topological
string, a rather daunting task.  In the B-model, one expects
Kodaira-Spencer theory to give rise to the gravitational story, but
in the twistor-string context this does not seem to play a role.

Berkovits' model~\cite{Berkovits:2004hg} is rather simpler: the
worldsheet path integral localizes on holomorphic (rather than
constant) maps, and worldsheet instantons of degree $d\geq1$ play
the role of the D1 branes in Witten's model. Berkovits' strings have
boundaries on a totally real (and hence Lagrangian) submanifold
$\bR\bP^{3|4}\subset\bC\bP^{3|4}$ which may be reminiscent of the
open A-model. However, spacetime Yang-Mills interactions arise not
from D branes wrapping $\bR\bP^{3|4}$, but via a worldsheet current
algebra, while gravitational modes are generated by vertex operators
on the same footing as those of Yang-Mills in the sense that both
are inserted on the worldsheet boundary. Moreover, $\bR\bP^3$
corresponds to a spacetime metric of signature $(++--)$ and it is
not clear that scattering theory makes sense in such a signature,
because the lightcone is connected and there appears to be no
consistent $\im\epsilon$ prescription.

In this paper we recast twistor-string theory as a heterotic string.
The first reason to suspect that a heterotic perspective is relevant
to the twistor-string is Nair's original
observation~\cite{Nair:1988bq} that Yang-Mills MHV amplitudes may be
obtained from a current algebra on a $\bP^1$ linearly embedded in
twistor space; such a current algebra arises naturally in a
heterotic model. Secondly, heterotic sigma models with complex
manifolds such as twistor space as a target automatically have (0,2)
worldsheet supersymmetry. This supersymmetry may be twisted so that
correlation functions of operators representing cohomology classes
of the scalar supercharge localize on holomorphic maps to twistor
space. So holomorphic curves in twistor space are naturally
incorporated as worldsheet instantons, as in Berkovits' model, and
no D-branes are necessary (or even possible). Thirdly, the twisted
theory depends only on the global complex structure of the target
$X$ and of a holomorphic bundle $E\to X$, as well as a certain
complex analytic cohomology class on $X$. At the perturbative level,
infinitesimal deformations of these structures correspond to
elements of the cohomology groups $H^1(X,T_X)$, $H^1(X,{\rm End} E)$
and $H^1(X,\Omega^2_{\rm cl})$, where $\Omega^2_{\rm cl}$ is the
sheaf of closed holomorphic 2-forms on $X$.  In the twistor context,
this dovetails very naturally with the Penrose transform which gives
an isomorphism between these cohomology groups (together with their
supersymmetric extensions) and the on-shell states of linearized
conformal supergravity and super Yang-Mills.  Thus the ingredients
of twistor-string theory combine very naturally in a heterotic
picture.

While our heterotic picture is closest in spirit to Witten's model, in
particular representing target space cohomology groups via Dolbeault
cohomology, twisted (0,2) models have recently been understood to be
very close cousins of $\beta\gamma$-systems through a quantum field
theoretic version of the {\v C}ech-Dolbeault isomorphism (see
\cite{Witten:2005px}, a paper that provided much of the stimulus for
this one). This relationship provides the link between the heterotic
and Berkovits' twistor-strings, with the latter becoming freed from
its dependence on split signature spactime.  It might be thought that
the connection to Witten's B-model plus D1-instantons might be taken
to be that the heterotic model provides the detailed theory of the
D1-instantons, but one then discovers that the open strings of the
$B$-model are redundant, and their corresponding degrees of freedom
and interactions are alreaded incorporated in the degree zero sector
of the heterotic string.

The paper is structured as follows.  In section~\ref{basicssec} we
review the theory of twisted (0,2) sigma models.  In
section~\ref{twistorsec}, we introduce the twistor-string model that
we will study. The target space of our model is (a region in) the
non-supersymmetric twistor space $\bP^3$, but we also include
fermions which are worldsheet scalars with values in a non-trivial
vector bundle $\cV\to\bP^3$. The fact that these fermions are
worldheet scalars means that vertex operstors can have arbitrary
dependence on them and so they play the role of the anti-commuting
coordinates on supertwistor space $\bP^{3|4}$. In this section we
show that the sigma model anomalies cancel, and study the moduli
space of worldsheet instantons.  In section~\ref{vertexsec} we
introduce the basic vertex operators of the model, paying particular
attention to those which correspond to deformations of the complex
structure or a NS $B$-field on the twistor space. These correspond
on spacetime to the conformal supergravity degrees of freedom.  In
section~\ref{ymsec} we introduce a further fermions (now spinors on
the worldsheet) with values in another bundle $E\to\bP^3$, and these
provide a coupling to Yang-Mills fields on spacetime.  In
section~\ref{stringsec} we promote the previously studied sigma
models to a string theory by coupling in a `$bc$ system', and study
the associated conformal anomaly.  In section~\ref{googlysec} we
give a more detailed discussion of the deformed supertwistor spaces,
in particular discussing the way in which the googly data is
encoded.  In section~\ref{relationsec} we show how this model
relates to both the Berkovits model and the original Witten model, in particular clarifying the role of the D1-D1 strings in Witten's picture.
In section~\ref{sftsec} we discuss the string field theory of the
disconnected prescription and derive the corresponding twistor
action.  We conclude with a discussion in section~\ref{discusssec}.

\section{A review of the twisted (0,2) sigma model}
\label{basicssec}

Let us begin by briefly reviewing the construction of a (0,2) non-linear
sigma model describing maps $\phi:\Sigma\to X$ from a compact Riemann
surface $\Sigma$ to a complex manifold $X$ (see
also~\cite{Witten:2005px,Adams:2005tc} for recent
work in a similar context). The basic fields in the model are
worldsheet scalars $\phi$, representing the pullback to $\Sigma$ of
coordinates on a local patch of $X$. Twisted (0,2) supersymmetry
requires that we pick a complex structure on $\Sigma$ and introduce
fields
\begin{equation}
\rho^i\in\Gamma(\Sigma,\ov K\otimes \phi^*T_X)
\qquad\qquad\qquad
\bar\rho^{\bar\jmath}\in\Gamma(\Sigma,\phi^*\ov{T}_X)
\label{tangent}
\end{equation}
where $\ov K$ is the anticanonical bundle on $\Sigma$ and $T_X$ is
the holomorphic tangent bundle on $X$. These fields are related to
the $\phi$s by the supersymmetry transformations
\begin{equation}
\begin{aligned}
&\delta \phi^i=\epsilon_2\rho^i&\qquad\qquad\qquad&
\delta\phi^{\bar\jmath}=\epsilon_1\bar\rho^{\bar\jmath}\\
&\delta \rho^i=\epsilon_1\delbar\phi^i
&&\delta\bar\rho^{\bar\jmath}=\epsilon_2\delbar\phi^{\bar\jmath}
\end{aligned}
\label{trans1}
\end{equation}
where $\epsilon_i$ are constant anticommuting parameters with
$\epsilon_1$ a scalar and $\epsilon_2$ a section of $\ov T_\Sigma$.
The transformation parameterized by $\epsilon_1$ may be defined
globally on $\Sigma$, whilst constant antiholomorphic vector fields
only exist locally on $\Sigma$ (except at genus 1), so $\epsilon_2$
may only be defined within a local patch on $\Sigma$, with
coordinates $(z,\bar z)$. Let these transformations be generated by
supercharges $\ov Q$ and $\ov Q^\dagger$, so that for a generic
field $\Phi$
\begin{equation}
\delta\Phi = \left[\epsilon_1\ov Q + \epsilon_2\ov Q^\dagger, \Phi\right].
\label{delta}
\end{equation}
with $\ov Q$ a scalar operator. It is straightforward to check that
$\ov Q^2=0$ and, on our local patch, also $(\ov Q^\dagger)^2=0$ and
$\{\ov Q,\ov Q^\dagger\}=\delbar$. These relations characterize
(0,2) (twisted) supersymmetry.

To write an action we pick a Hermitian metric $g$ on $X$. The basic
action for a non-linear sigma model is then
\begin{equation}
\begin{aligned}
S_1&=\int_\Sigma\!|\rd^2z|\ \frac{1}{2}g_{i\bar\jmath}\left(\del_{\bar
z}\phi^i\del_z\phi^{\bar\jmath}
+ \del_z\phi^i\del_{\bar z}\phi^{\bar\jmath}\right)-\rho^i_{\bar z}\nabla_z\bar\rho^{\,\bar\jmath}\\
&=\left\{\ov Q,\int_\Sigma\!|\rd^2z|\  g_{i\bar\jmath}\rho^i_{\bar
z}\del_z\phi^{\bar\jmath}\right\}+\int_\Sigma\phi^*\omega
\label{act}
\end{aligned}
\end{equation}
where $\nabla : \Gamma(\Sigma,\phi^*\ov
T_X)\to\Gamma(\Sigma,K\otimes\phi^*\ov T_X)$ is the pullback to
$\Sigma$ of the Hermitian connection on $\ov T_X$ and $\omega = \im
g_{i\bar\jmath}\,\rd\phi^i\wedge\rd\phi^{\bar\jmath}$. If
$\rd\omega=0$ so that $X$ is K\"ahler, the action is invariant under
the (0,2) transformations \ref{trans1} and the connection $\nabla$
is Levi-Civita. Because the action is $\ov Q$-exact upto the
topological term $\int_\Sigma\phi^*\omega$, correlation functions of
operators in the $\ov Q$-cohomology will not depend on the choice of
Hermitian metric $g$. They do depend on the K\"ahler class of
$\omega$ together with the complex structures on $X$ and $\Sigma$,
which were used to define the transformations~\ref{trans1}.

There are various generalizations beyond this basic
picture~\cite{Gates:1984nk,Hull:1985jv,Witten:2005px}. Firstly, by
introducing a $\del$-closed (2,0) form $t$ we may deform \ref{act}
by
\begin{equation}
\begin{aligned}
\delta S_1&=\im \int_\Sigma\!|\rd^2z|\ \del_{\bar k}t_{ij}\bar\rho^{\bar k}\rho^i_{\bar z}\del_z\phi^j
+ t_{ij}\del_{\bar z}\phi^i\del_z\phi^j\\
&=\im\left\{\ov Q,
\int_\Sigma\!|\rd^2z|\ t_{ij}\rho^i_{\bar z}\del_z\phi^j\right\} \ .
\end{aligned}
\label{bdefaction}
\end{equation}
If $t$ is globally defined on $X$, then this deformation is $\ov
Q$-trivial and $t$ does not affect correlators of operators
representing $\ov Q$-cohomology classes. More interesting is the
case where $t$ is defined only on the local patches of some cover
$\{U_\alpha\}$ of $X$, where $\alpha$ indexes the cover. If the
differences $t^{(\alpha)}-t^{(\beta)}$ are holomorphic on each
overlap $U_\alpha\cap U_\beta$, then they piece together to form an
element $\cH$ of the cohomology group $H^{0,1}(X,\Omega^{2,0}_{\rm
cl})$ where $\Omega^{2,0}_{\rm cl}$ is the sheaf of $\del$-closed
(2,0)-forms on $X$. The correlation functions are then sensitive to
this class. We can also think of $\cH$ in terms of a Dolbeault
representative, a global $(2,1)$-form satisfying
$\del\cH=\delbar\cH=0$ obtained as $\cH=\delbar t^\alpha$.  Whilst
the second line of~\ref{bdefaction} makes it clear that this
modification is invariant under $\ov Q$ transformations, $\delta
S_1$ is invariant under the full (0,2) supersymmetry if and only if
$\cH$ satisfies $\cH=2\im\del\omega$. Correspondingly, in the
presence of $\cH$ the hermitian metric connection $\nabla$ has
torsion $T^i_{\ jk}=g^{i\bar n}\cH_{\bar n jk}$.

Hull and Witten~\cite{Hull:1985jv,Witten:2005px} observed that
locally this geometric structure can be derived from a smooth 1-form
$K(\phi,\bar\phi)$ which serves as a potential for both $t$ and
$\omega$ by $\im t=2\del K$ and $\omega= 2\Re\,\delbar K$ (and so
also $\cH=\del\delbar K$).  The action is then given by
\begin{eqnarray}
S_1&=&\int |\rd^2 z|\left(
K_{i,\bar\jmath}\del_{\bar z}\phi^{\bar\jmath}\del_z\phi^i
+\ov K_{\bar\imath,j}\del_{\bar z}\phi^j\del_z\phi^{\bar\imath}\right.
\nonumber\\
& &\qquad \left.
-(K_{i,\bar\jmath}\bar\rho^{\bar\jmath}\del_z\rho^i
+\ov K_{\bar\imath,j}\rho^j_{\bar z}\del_z\bar\rho^{\bar\imath})
+(K_{i,\bar\jmath k}\bar\rho^{\bar\jmath}\rho^k_{\bar z}\del_z\phi^i
-\ov K_{\bar\imath,j\bar l}\rho^j_{\bar z}\bar\rho^{\bar\imath}
\del_z\phi^{\bar\imath})
\right)\label{wittenact} \\
&=&
\left\{\ov Q,\int |\rd^2 z | \left((K_{i,\bar\jmath}+\ov
K_{\bar\jmath,i})\rho^i_{\bar z} \del_z\phi^{\bar\jmath}
-(K_{i,j}-K_{j,i})\rho^i_{\bar z}\del_z\phi^j\right)\right\}\ .
\nonumber
\end{eqnarray}
It will also be useful to introduce a $(1,1)$-form $b$ as $b=\delbar K$.
Then $b=B + \im\omega$ where $B$ is the usual $B$-field of
string theory and $\cH=\del b$.  See~\cite{Witten:2005px} for a fuller
discussion of the geometry underlying these models.

The most important feature of twisted (0,2) models is that the
action is $\ov Q$-exact (except for topological terms) so the path
integral localizes on $\ov Q$-invariant solutions to the equations
of motion. In particular, the transformation $\{\ov Q,\rho^i_{\bar z}\}=\del_{\bar z}\phi^i$ shows that such invariant configurations
are holomorphic maps, or worldsheet instantons.  The full action
evaluated on such invariant solutions is $\int_\Sigma \phi^* b$. If
$b$ is not globally defined, one can only make sense of this
expression provided the underlying de Rham cohomology class of $\cH$
is integral.

\subsection{Coupling to bundles}

We can also incorporate holomorphic bundles over $X$: let $\cV\to X$
be a holomorphic vector bundle and introduce fields
\begin{equation}
\begin{aligned}
&\psi^a\in\Gamma(\Sigma,K^s\otimes\phi^*\cV)
&\quad\qquad
&\bar\psi_a\in\Gamma(\Sigma,K^{1-s}\otimes\phi^*\cV^\vee)\\
&r^a\in\Gamma(\Sigma,\ov K\otimes K^s\otimes\phi^*\cV)&
&\bar r_a\in\Gamma(\Sigma,K^{1-s}\otimes\phi^*\cV^\vee)
\end{aligned}
\label{bunferms}
\end{equation}
where $\cV^\vee$ is the dual bundle to $\cV$. Note that classically,
twisted (0,2) supersymmetry does not fix the spin of these
left-moving fields and at present we allow them to be sections of
$K^s$ for any half-integer $s$. For what follows, it will be
convenient to choose the fields in \ref{bunferms} to behave
equivariantly under $\ov Q$ transformations and gauge
transformations on $\cV$ (as in~\cite{Adams:2005tc}), obtaining
\begin{equation}
\begin{aligned}
&\delta\psi^a = \epsilon_2(r^a + A_{i\ b}^{\ a}\psi^b\rho^i)
&\quad\quad
&\delta\bar\psi_a = \epsilon_1\bar r_a\\
&\delta r^a = \epsilon_1(\ov D\psi^a + F_{i\bar\jmath\,\ b}^{\,\ a}\psi^b\rho^i\bar\rho^{\bar\jmath})
+\epsilon_2A_{i\ b}^{\ a}r^b\rho^i&
&\delta\bar r_a = \epsilon_2\delbar\bar\psi_a
\end{aligned}
\label{buntrans}
\end{equation}
where $\ov D :
\Gamma(\Sigma,K^s\otimes\phi^*\cV)\to\Gamma(\Sigma,\ov K\otimes
K^s\otimes\phi^*\cV)$ is a connection on $K^s\otimes\phi^*\cV$. One
can check that the (0,2) algebra is satisfied provided $\cV$ is
holomorphic so that $F_{ij}=F_{\bar\imath\bar\jmath}=0$. The action
for these bundle-valued fields is taken to be
\begin{equation}
\begin{aligned}
S_2&=\int_\Sigma\!|\rd^2z|\
\bar\psi_a D_{\bar z}\psi^a
+F_{i\bar\jmath\,\ b}^{\,\ a}\bar\psi_a\psi^b\rho^i_{\bar z}\bar\rho^{\bar\jmath} + \bar r_ar^a\\
&=\left\{\,\ov Q\,,\int_\Sigma\!|\rd^2z|\, \bar\psi_a r^a_{\bar z}\right\} \ .
\end{aligned}
\label{act2}
\end{equation}
In particular, this shows that $r$ and $\bar r$ are auxiliary and decouple.

Classically, the stress-energy of $S_1+S_2$ has non-vanishing components
\begin{equation}
\begin{aligned}
T_{zz} &=g_{i\bar\jmath}\del_z\phi^i\del_z\phi^{\bar\jmath}
+ \bar\psi_aD_z\psi^a\\
T_{\bar z\bar z}&=g_{i\bar\jmath}\left(\del_{\bar z}\phi^i\del_{\bar z}\phi^{\bar\jmath}
+\rho^i_{\bar z}\nabla_{\bar z}\bar\rho^{\bar\jmath}\right)
=\left\{\ov Q,g_{i\bar\jmath}\rho^i_{\bar z}\del_{\bar z}\phi^{\bar\jmath}\right\}\ .
\label{T}
\end{aligned}
\end{equation}
Since $T_{\bar z\bar z}=\{\ov Q,\,\cdot\,\}$, as discussed
in~\cite{Witten:2005px} all the Laurent coefficients $\bar L_n$ of
$T_{\bar z\bar z}$ are also $\ov Q$-exact. In particular, $\bar L_0
= \{\ov Q,\ov G_0\}$ for some $\ov G_0$, so that $\bar L_0$ maps
$\ov Q$-closed states to $\ov Q$-exact ones and is thus zero in
cohomology. But for any state of antiholomorphic weight $\bar h\neq
0$, $\bar L_0/\bar h$ is the identity, so the $\ov Q$-cohomology
vanishes except at $\bar h=0$. Furthermore, the fact that $T_{\bar
z\bar z}$ is $\ov Q$-exact means that correlation functions
$\langle\prod_{i=1}^n\cO_i(z_i)\rangle$ of $\ov Q$-closed operators
depend only holomorphically on the insertion points $\{z_i\}\in\Sigma$.
Were we studying a model with twisted (2,2) supersymmetry, exactly
the same argument for the left-movers would lead us to conclude that
operators in the BRST cohomology must also have $h=0$, and that
correlation functions are actually independent of the insertion
points. However, here $T_{zz}\neq\{\ov Q,\ldots\}$ and so there is
an infinite tower of $\ov Q$-cohomology classes depending on
$h\in\bZ_{\geq0}$, and the twisted (0,2) model is a conformal,
rather than topological, field theory.

If we choose $\cV\cong T_X$ and set $s=0$ the total action $S_1+S_2$
in fact has twisted (2,2) worldsheet supersymmetry and is the action
of the A-model, while choosing $\cV=T_X$ but keeping $s=1/2$ gives a
half-twisted version of this (2,2) theory. (0,2) models allow for
more general choices of $\cV$, as is familiar from compactifications
of the physical heterotic string where $\cV$ is a subbundle of the
$E_8\times E_8$ or $Spin(32)/\bZ_2$ gauge bundles in ten dimensions
(where, in the physical string, $s=1/2$). In that context, setting
$\cV=T_X$ corresponds to the `standard embedding' of the gauge
connection in the spin connection of the compactification manifold.
For recent work on twisted (0,2) models related to heterotic
compactification,
see~\cite{Basu:2003bq,Adams:2003zy,Beasley:2003fx,Katz:2004nn,Adams:2005tc}.

\section{The twistor target space}
\label{twistorsec}

In this paper, we will reformulate twistor-string theory as a (0,2)
model. One might anticipate that we should take $X$ to be a region
in $\bP^{3|4}$ as in~\cite{Witten:2003nn,Berkovits:2004hg} but,
while this may well be a reasonable way to proceed, in its most
na\"\i ve form a (0,2) model with $\bP^{3|4}$ target leads to
difficulties both in understanding the role of the bosonic
worldsheet superpartners of the fermionic directions, and in
handling the antiholomorphic fermionic directions without the
possibility of appealing to a `D-brane at $\bar\psi=0$', since
heterotic models do not possess D-branes.

We therefore adopt a different strategy in which the basic target
space is $\bP^3$, the non-supersymmetric, projective twistor space
of flat spacetime. The fermionic directions of $\bP^{3|4}$ are
incorporated by coupling to a bundle $\cV\equiv\cO(1)^{\oplus4}$ as
in \ref{bunferms}-\ref{act2} with $s=0$. With this choice of $s$,
the $\psi^a$ are anticommuting worldsheet scalars and so provide the
fields that were used in the original twistor-string
theories~\cite{Witten:2003nn,Berkovits:2004hg} to describe
holomorphic coordinates on the fermionic directions of $\bP^{3|4}$.
The vertex operators will be seen to correspond to perturbations of
both the complex structure and of the NS flux $\cH$, and these
perturbations can also have arbitrary dependence on $\psi^a$.  With
$s=0$, $\bar\psi_a$ are sections of
$K\otimes\phi^*\left(\cO(1)^{\oplus4}\right)^\vee)$ and are thus
worldsheet (1,0) forms, so $\psi$ and $\bar\psi$ are naturally on a
different footing. Correspondingly, we will see that the dependence
of the vertex operators on $\bar\psi^a$ can be at most linear. Thus
our model is equivalent to working on a $\bP^{3|4}$ target, at least
at the linearized level determined by the vertex operators. In order
to incorporate Yang-Mills, in section~\ref{ymsec} we will also
couple to a bundle with action~\ref{act2}, but where $s=1/2$.  In
this case the allowed vertex operators are different and will
correspond to twistor data for super Yang-Mills fields.

Initially,  to consider the quantum theory we will take the action
to be $S=S_1+S_2$ as in~\ref{act} \&~\ref{act2}, with target
$\bP^3-\bP^1$ and bundle $\cV=\cO(1)^{\oplus 4}$ with associated
fermions $\psi^a\in\Gamma(\Sigma,\phi^*\cV)$ and
$\bar\psi_{az}\in\Gamma(\Sigma,K\otimes\phi^*\cV^\vee)$. The
K{\"a}hler structure is given by the Fubini-Study metric which
induces a metric and compatible connection on $\cO(1)$.  We postpone
the coupling to Yang-Mills until section~\ref{ymsec}. Note that the
first-order action for the $\psi\bar\psi$-system is reminiscent of
Berkovits' model~\cite{Berkovits:2004hg}; we will make the
relationship more precise in section~\ref{berkovitssec}.

\subsection{Anomalies}
\label{anomsec}

With these choices of $X$, $\cV$ and $s$ we must show that the
classical  action $S_1+S_2$ of equations~\ref{act} \&~\ref{act2}
defines a sensible quantum theory.

\subsubsection{Sigma model anomalies}
\label{siganomsec}
Field theories containing chiral fermions may fail to define a
quantum theory because of the presence of sigma model anomalies:
integrating out the fermions gives a one-loop determinant which must
be treated as a function of the bosonic fields. However this
determinant is really a section of a line bundle  $\cL\to{\rm
Maps}(\Sigma,X)$ over the space of maps and we can only make a
canonical identification of this section with a function if the
determinant line bundle is flat~\cite{Moore:1984ws}. In twisted
(0,2) models, integrating out the non-zero-modes of $\rho$ and
$\psi$ gives a factor $\det'\nabla\,\det'\ov D$ which depends on the
map $\phi$ through the pullback of $\ov T_X$ in $\nabla$ and the
pullback of $\cV$ in $\ov D$. Since $\det'\nabla = \det'\triangle
/\det'\delbar_{\phi^*T_X}$ and the $\zeta$-regularized determinant
of the self-adjoint Laplacian $\triangle$ is always well-defined,
the anomaly is governed by the virtual bundle $\cV\ominus T_X$.

The geometric index theorem of Bismut and
Freed~\cite{Bismut:1986ab,Freed:1986zx} states that the curvature of
the Quillen connection~\cite{Quillen} on $\cL$ is given by
\begin{equation}
\begin{aligned}
F^{(\cL)}&=\int_\Sigma\left.{\rm Td}(T_\Sigma)\phi^*{\rm ch}(\cV\ominus T_X)\right|_{(4)}\\
&=\int_{\Sigma}\frac{{\rm c}_1(T_\Sigma)}{2}\phi^*\left({\rm c}_1(\cV)-{\rm
c}_1(T_X)\right)+ \int_{\Sigma}\phi^*\left({\rm ch}_2(\cV)-{\rm ch}_2(T_X)\right)\ .
\end{aligned}
\label{sigmaanom}
\end{equation}
The first term in \ref{sigmaanom} is not present in the physical
heterotic string and arises here because the worldsheet fermions
$\rho$, $\psi$ and their duals are scalars and 1-forms. This term
depends on the genus of $\Sigma$ and so it must vanish separately if
the sigma model is to be well-defined on an arbitrary genus
worldsheet. Requiring that the second term also vanishes is then
familiar as a consistency condition for the Green-Schwarz
mechanism\footnote{On $\bP^3$, the background Neveu-Schwarz
fieldstrength $H$ vanishes, so the left hand side of \ref{GS} is
zero as a form, and not just in cohomology. Consequently the Quillen
connection must be flat, rather than merely have vanishing first
Chern class, and so $\Omega^{(\cL)}$ itself must vanish. For target
spaces with torsion, a flat connection on $\cL$ may be constructed
by modifying the Quillen connection by a term involving
$H$~\cite{Freed:1986zx}.}
\begin{equation}
dH = {\rm ch}_2(T_X)-{\rm ch}_2(\cV)\ .
\label{GS}
\end{equation}
When $\cV=T_X$ as in the A-model, $F^{(\cL)}$ vanishes
trivially. In the B-model, $\cV=T_X^\vee$ so $F^{(\cL)}=0$ if and
only if ${\rm c}_1(T_X)=0$. For more general (0,2) models, the
condition that \ref{sigmaanom} should vanish highly constrains the
admissible choices of $\cV$.

In the twistor-string case at hand, $X=\bP^3$ and $\cV=\cO(1)^{\oplus4}$.
The bundle $\cO(1)^{\oplus4}$ appears in the Euler sequence
\begin{equation}
0\to\cO\to\cO(1)^{\oplus4}\to T_{\bP^3}\to0
\label{Euler}
\end{equation}
in which the first map is multiplication by the homogeneous
coordinates $Z^\alpha$ on $\bP^3$, and the second map is
$V^\alpha\to V^\alpha\del/\del Z^\alpha$ which defines the tangent
bundle of projective space as a quotient of that on the
non-projective space. Since \ref{Euler} is exact,
\begin{equation}
{\rm c}(\cO(1)^{\oplus4})={\rm c}(\cO)\,{\rm c}(T_{\bP^3})
={\rm c}(T_{\bP^3})
\label{Chern}
\end{equation}
so {\it all} the Chern classes of $\cO(1)^{\oplus4}$ agree with
those of $T_{\bP^3}$, ensuring that \ref{sigmaanom} vanishes. By
comparison, for $\bP^{3|4}$ the Euler sequence reads
\begin{equation}
0\to\cO\to\bC^{4|4}\times\cO(1)\to T_{\bP^{3|4}}\to 0
\label{superEuler}
\end{equation}
so that
\begin{equation}
{\rm ch}(T_{\bP^{3|4}})= {\rm ch}(\bC^{4|4}\times\cO(1)) - {\rm ch}(\cO)
={\rm sdim}\,\bC^{4|4}\ {\rm ch}(\cO(1)) -1 = -1
\label{superChern}
\end{equation}
showing that (formally) ${\rm sdim} \bP^{3|4}=-1$ while all its Chern
classes vanish. Note in particular that triviality of the Berezinian
of $\bP^{3|4}$ is equivalent to the statement that $K_{\bP^3}\simeq
\bigwedge^{\rm top}(\cO(1)^{\oplus4})^\vee$, while ${\rm
sdim}\,\bP^{3|4}=-1$ is equivalent to the fact that the vanishing
locus of a generic section of $\cO(1)^{\oplus4}$ has virtual
dimension $-1$. We now wish to show that a similar relationship
holds at the level of the instanton moduli space.

\subsubsection{Anomalous symmetries and the instanton moduli space}
\label{globanomsec}

The action $S_1+S_2$ is invariant under a global $U(1)_F\times U(1)_R$
symmetry, where $U(1)_R$ is the automorphism group of the (0,2)
superalgebra and $U(1)_F$ is a left-moving `flavour symmetry'
associated to the bundle-valued fermions. As in~\cite{Adams:2005tc},
we take $\rho$ and $\bar\rho$ to have respective charges $(0,-1)$
and $(0,1)$ under $U(1)_F\times U(1)_R$, while $\psi$ and $\bar\psi$
have charges $(1,0)$ and $(-1,0)$; $\phi$ is uncharged. These
symmetries are violated by the path integral measure because the
fermion kinetic operators have non-zero index. The violation is tied
directly to the geometry of the instanton moduli space and restricts
the combinations of vertex operators that can contribute to a
non-vanishing amplitude.

The anomalies arise from the index theorem applied to the fermion
kinetic terms. The kinetic term $g_{i\bar\jmath}\rho^i_{\bar
z}\nabla_z\bar\rho^{\bar\jmath}$ implies that a $\bar\rho$ zero-mode
is an {\it antiholomorphic} section of $\phi^*\ov T_{\bP^3}$ and so
is complex conjugate to an element of $H^0(\Sigma,\phi^*T_{\bP^3})$.
Similarly, zero-modes of $g_{i\bar\jmath}\rho^i_{\bar z}$ are
complex conjugate to elements of
$H^0(\Sigma,K\otimes\phi^*T_{\bP^3}^\vee)\simeq
H^1(\Sigma,\phi^*T_{\bP^3})$, by Serre duality.  The
Hirzebruch-Riemann-Roch theorem then says that the difference in the
complex dimensions of the spaces of such zero-modes on a worldsheet
of genus $g$ is
\begin{equation}
\begin{aligned}
h^0(\Sigma,\phi^*T_{\bP^3}) - h^1(\Sigma,\phi^*T_{\bP^3})
&=\int_{\Sigma}\phi^*{\rm c}_1(T_{\bP^3})+{\rm dim}(\bP^3)
\frac{{\rm c}_1(T_\Sigma)}{2}\\
&=4d+3(1-g) \label{Ranom}
\end{aligned}
\end{equation}
for a degree $d$ map to twistor space.

Given a holomorphic map $\phi$, a nearby map $\phi+\delta\phi$ is
also holomorphic provided $\delta\phi\in H^0(\Sigma,\phi^*T_X)$.
Consequently, the holomorphic tangent bundle $T_\cM$ to instanton
moduli space $\cM$ has fibre $T_\cM|_\phi=H^0(\Sigma,\phi^*T_X)$.
The $\bar\rho$ zero-modes are anticommuting elements of
$\ov{H^0(\Sigma,\phi^*T_X)}$ and thus represent (0,1)-forms on
$\cM$. Maps $\phi$ at which $h^1(\Sigma,\phi^*T_X)=0$ are
non-singular points of the instanton moduli space and the tangent
space there has dimension equal to the above index. In the
twistor-string case, either at genus zero or when the degree is
sufficiently larger than the genus, such points form a dense open
set of the instanton moduli space. So our model has no $\rho^i$
zero-modes and $4d+3$ $\bar\rho^{\bar\jmath}$ zero-modes at genus
zero. In the rational case with target $\bP^3$, a degree $d$ map can
be expressed as a polynomial of degree $d$ in the homogeneous
coordinates $Z^\alpha$, as $Z^\alpha(\sigma) = \sum_{i=0}^d
A^\alpha_{\ i}\sigma^i$. The coefficients $A^\alpha_{\ i}$ are
therefore homogeneous coordinates on the moduli space $\cM$ and one
can identify\footnote{More accurately, the moduli
  space of instantons in the non-linear sigma model at genus zero is a
  dense open subset in $\bP^{3+4d}$, noncompact because of
  `bubbling'. A linear sigma model presentation provides a natural
  compactification~\cite{Morrison:1994fr} of $\cM$ to $\bP^{4d+3}$ and
  we will henceforth work over this compact moduli space.}
$\cM\cong\bP^{4d+3}$ for genus zero maps to $\bP^3$.

Turning now to the $\psi$ fields, the kinetic term $\bar\psi_a\ov
D\psi^a$ shows that a $\psi$ zero-mode represents an element of
$H^0(\Sigma,\phi^*\cV)$ while a $\bar\psi$ zero-mode represents an
element of $H^0(\Sigma,K\otimes\phi^*\cV^\vee)\cong
H^1(\Sigma,\phi^*\cV)^\vee$, again by Serre duality. Hence the
difference in the number of zero-modes is
\begin{equation}
\begin{aligned}
h^0(\Sigma,\phi^*\cV)-h^1(\Sigma,\phi^*\cV) &=\int_\Sigma\phi^*{\rm
c}_1(\cV)
+{\rm rk}(\cV)\frac{{\rm c}_1(T_\Sigma)}{2}\\
&=4(d+1-g),
\end{aligned}
\label{Fanom}
\end{equation}
for $\cV=\cO(1)^{\oplus4}$. This anomaly is familiar in the
twistor-string story. It says that correlation functions on a degree
$d$, genus $g$ curve vanish unless the path integral contains an
insertion of net $U(1)_F$ number $4(d+1-g)$. We will see that, just
as in the Witten and Berkovits twistor-strings, the vertex operators
naturally form spacetime $\cN=4$ multiplets by depending
polynomially on $\psi$, but not $\bar\psi$. In particular, a
correlation function involving $n$ external gluons of
positive\footnote{In our conventions, elements of the cohomology
group $H^1(\bT',\cO(-2h-2))$ correspond via the Penrose
transform to spacetime fields of helicity $h$, so that in particular
a negative helicity gluon corresponds to a twistor wavefunction of
weight $0$.} helicity and arbitrary gluons of negative helicity is
supported on a worldsheet instanton of degree
\begin{equation}
d=n - 1+g\ ,
\label{MHV}
\end{equation}
as in~\cite{Witten:2003nn}. More generally, scattering amplitudes of
$n_h$ external SYM states of helicity $h$ are supported on curves of
degree
\begin{equation}
d= g-1 + \sum_{h=-1}^1 \frac{h+1}{2}\, n_h
\label{MHV2}
\end{equation}
and must necessarily vanish unless $d\in \bZ_{\geq0}$.

As discussed by Katz \& Sharpe in~\cite{Katz:2004nn}, just as for
the $\bar\rho$ zero-modes, the $\psi$ zero-modes may be interpreted
geometrically in terms of a bundle (really, a sheaf) over $\cM$. Consider
the diagram
\begin{equation}
\begin{CD}
\cM\times\Sigma @>\Phi>> X \\
\ \, @V\pi VV\\
\,\cM
\end{CD}
\label{universal}
\end{equation}
where $\Phi$ is the universal instanton and $\pi$ the obvious
projection. Given a sheaf $\cV$ on $X$ we can construct a sheaf
$\cW$ over $\cM$ by pulling back $\cV$ to $\cM\times\Sigma$ via the
universal instanton, and then taking its direct image under the
projection map, {\it i.e.} $\cW\equiv\pi_*\Phi^*\cV$. The direct
image sheaf is defined so that its sections over an open set
$U\subset\cM$ are
\begin{equation}
\cW(U)=(\pi_*\Phi^*\cV)(U)
=(\Phi^*\cV)(\pi^{-1}U)= H^0(U\times\Sigma,\Phi^*\cV)\ ,
\label{Wdef}
\end{equation}
so that over a generic instanton, $\cW|_\phi=H^0(\Sigma,\phi^*\cV)$
with dimension $4(d+1-g)$. Consequently, we may generically
interpret a $\psi$ zero-mode as a point in the fibre $\cW|_\phi$.

For families of instantons for which there are no $\rho$ or
$\bar\psi$ zero-modes ({\it i.e.} whenever the higher direct image
sheaves $R^1\pi_*\Phi^*T_X$ and $R^1\pi_*\Phi^*\cV$ vanish), the
definition of $\cW$ shows that it has first Chern class~\cite{Katz:2004nn}
\begin{equation}
{\rm c}_1(\cW)=\int_\Sigma{\rm Td}(T_\Sigma)\,\Phi^*{\rm ch}(\cV)|_{(4)}
\end{equation}
so the condition ${\rm ch}(\cV)={\rm ch}(T_X)$ ensures that
${\rm c}_1(\cW)={\rm c}_1(T_\cM)$, or
\begin{equation}
\bigwedge\!\!^{\rm top}\,\cW^\vee\simeq K_\cM\ .
\label{Wtop}
\end{equation}
This isomorphism is important in computing correlation functions:
operationally, to integrate out the $\psi$ zero-modes one merely
extracts the coefficient of the $\psi$s in the vertex operators,
restricting ones attention to instantons whose degree is determined
by \ref{MHV2}. This coefficient is a section of $\bigwedge^{\rm
top}\cW^\vee$, so by \ref{Wtop} we may interpret it as a top
holomorphic form on instanton moduli space.

Again, this story has a familiar counterpart in the original
construction of twistor-strings~\cite{Witten:2003nn} as a theory
with target space $\bP^{3|4}$. Assuming that there is a dense open
subset of the moduli space over which there are no $\bar\psi$
zero-modes, \ref{Wtop} shows that the total space of the bundle
$\cW$, parity reversed on the fibres, can be thought of as a
Calabi-Yau supermanifold with a canonically\footnote{The holomorphic
volume form is defined upto scale, as is the
isomorphism~\ref{Wtop}.} defined holomorphic volume form (or
Berezinian). In particular, at genus zero there are no $\bar\psi$ or
$\rho$ zero-modes, and \ref{Wtop} simply states the isomorphism
$K_{\bP^{4d+3}}\simeq\cO(-4-4d)$. This is the (0,2) analogue of the
statement that the moduli space of rational maps to $\bP^{3|4}$ is
the supermanifold $\bP^{4d+3|4d+4}$ with trivial Berezinian.

Beyond genus zero, there can be zero-modes of both $\rho$ and
$\bar\psi$, and the dimension of $\cM$ and rank of $\cW$ may jump as
we move around in instanton moduli space. However, the indices
\ref{Ranom} \& \ref{Fanom} remain constant and so the selection rule
\ref{MHV2} is not affected by such excess zero-modes. To obtain
non-zero correlation functions we must now expand the action in
powers of the four-fermi term $F_{i\bar\jmath\,\ b}^{\,\ a}\bar\psi_a\psi^b\rho^i\bar\rho^{\bar\jmath}$ until the excess
zero-modes are soaked up. This is analogous to the way
(2,2) models construct the Euler class of the obstruction
sheaf~\cite{Aspinwall:1991ce}, but (0,2) models have the added
complication that $h^1(\Sigma,\phi^*T_X)$ may not equal
$h^1(\Sigma,\phi^*\cV)$, so that it may be necessary to absorb some
of the factors of $\rho\bar\rho$ or $\bar\psi\psi$ using their
respective propagators~\cite{Katz:2004nn}. Generically, when $d$ is
much larger than $g$ there are no excess zero-modes and \ref{Wtop}
again tells us that the moduli space of instantons from a fixed
worldsheet behaves as a Calabi-Yau supermanifold\footnote{See also
work by Movshev~\cite{Movshev:2006py}.}.

Incidentally, had we started with an untwisted model involving
worldsheet fermions that are sections of the square roots of the
canonical or anticanonical bundles, the anomaly in both the $U(1)_F$
and $U(1)_R$ symmetries would be $4d$, independent of the genus. A
diagonal subgroup of $U(1)_F\times U(1)_R$ would be anomaly free and
could be used to twist the spins of the fermions. One might compare
this to a (2,2) model on a K\"ahler manifold. There, a diagonal
subgroup of the $U(1)\times U(1)$ $R$-symmetry group is guaranteed
to be anomaly free simply because the left- and right-moving
fermions take values in the same bundle. Twisting by this subgroup
leads to the A-model. Even though the left- and right-moving
fermions of our (0,2) model are valued in different bundles, the
same subgroup is still anomaly free, again because of \ref{Chern}.

\subsection{Worldsheet perturbative corrections}
\label{betasec}

Because $T_{zz}\neq\{\ov Q,\cdot\}$, twisted (0,2) models are
conformal rather than topological field theories and we must examine
the effect of worldsheet perturbative corrections on the $\ov
Q$-cohomology. (0,2) supersymmetry ensures\footnote{In terms of
superfields, the most general action with (0,2) supersymmetry may be
written as $\int\!\rd^2\bar\theta\,\cD + \int\!\rd\bar\theta\,
\Gamma + \int\!\rd\bar\theta^\dagger\,\Gamma'$. The first two terms
are $\ov Q$-exact, while the third is not generated by quantum
corrections if it is not present at the classical level.} that
quantum corrections to the action will always be of the form $\{\ov
Q,\int_\Sigma \ldots\}$ so $T_{\bar z\bar z}$ will always remain
$\ov Q$-exact. Likewise~\cite{Witten:2005px,Adams:2005tc}, although
quantum corrections may lead to a violation of scale invariance,
since $T_{z \bar z}$ has antiholomorphic weight $\bar h=1$, any such
violation is always $\ov Q$-exact and worldsheet perturbative
corrections will not affect correlators representing $\ov
Q$-cohomology classes. One-loop corrections to worldsheet instantons
also have the effect of modifying the classical weighting by
$\int_\Sigma\phi^*\omega$ by the pullbacks of ${\rm c}_1(T_X)$ and
${\rm c}_1(\cV)$~\cite{Alvarez-Gaume:1985ww,Callan:1985ia}; these
corrections cancel in the twistor-string.

The only remaining issue is the correction to $T_{zz}$. Classically,
as in equation~\ref{T} we have
\begin{equation}
T_{zz}=g_{i\bar\jmath}\del_z\phi^i\del_z\phi^{\bar\jmath}+\bar\psi_{a\,z}D_z\psi^a \,
\label{holT}
\end{equation}
which is not $\ov Q$-exact, and obeys $\{\ov Q, T_{zz}\}=0$ only
once one enforces the $\rho$ equation of motion and vanishing of the
auxiliary fields $r$. Consequently, loop corrections to the
worldsheet effective action can easily upset $\ov Q$-closure of
$T_{zz}$. At 1-loop, the action receives a correction
\begin{equation}
\Delta S^{\rm 1\,loop} \propto\left\{\ov Q,\int_\Sigma\!|\rd^2z|\
R_{i\bar\jmath}\rho^i_{\bar z}\del_z\phi^{\bar\jmath} +
g^{i\bar\jmath}F_{i\bar\jmath\,\ b}^{\,\ a}\bar\psi_{a\,z} r^b_{\bar z}\right\}
\label{effact}
\end{equation}
and generically $T^{\rm 1\,loop}_{zz}$ is not $\ov Q$-closed unless
the target metric is Ricci-flat and the background connection on
$\cV$ obeys the Hermitian Yang-Mills equations so that this
correction vanishes. Neither of these conditions hold when
$X\cong\bP^3$ and $\cV\cong\cO(1)^{\oplus4}$. However, if $g$ is the
Fubini-Study metric then $SU(4)$ symmetry constrains
$R_{i\bar\jmath}=4g_{i\bar\jmath}$, while the curvature of
$\cO_{\bP^3}(1)^{\oplus4}$ obeys $F_{i\bar\jmath\,\ b}^{\,\
a}=g_{i\bar\jmath}\delta^a_{\ b}$ so that the 1-loop
correction~\ref{effact} is proportional to the classical action.
Consequently, the field equations are unaltered and $\{\ov Q, T^{\rm
1\,loop}_{zz}\}=0$ still holds. Similar results presumably hold for
higher loops in the worldsheet theory.

In a model with $\bP^{3|4}$ target space, these issues are more
straightforward: since ${\rm c}_1(T_{\bP^{3|4}})=0$ one can find a
Ricci-flat metric (the Fubini-Study metric on the
superspace~\cite{Witten:2003nn}) in which all one-loop corrections
vanish and there is always a metric in the same K\"ahler class
in which loop corrections vanish to any order. We have not taken
this route for the reasons discussed previously.

\section{Vertex operators and (0,2) moduli}
\label{vertexsec}

We now wish to determine the vertex operators representing $\ov
Q$-cohomology classes. Since the action is $\ov Q$-exact (upto the
topological term), correlation functions of such operators localize
on a first-order neighbourhood of the instanton moduli space
$\cM\subset{\rm Maps}(\Sigma,X)$, just as for the A-model.
Consequently, the one-loop approximation is exact for directions
normal to $\cM$ in ${\rm Maps}(\Sigma,X)$.

To construct these vertex
operators~\cite{Witten:2005px,Adams:2005tc}, we first note that they
must all be independent of $\rho^i_{\bar z}$, since this field has
antiholomorphic weight 1 (and the (0,2) theory does not contain any
fields with $\bar h <0$). Similarly, they must be independent of
antiholomorphic worldsheet derivatives of any of the fields.
However, (0,2) supersymmetry does not impose any constraints on the
holomorphic conformal weight, so {\it a priori} vertex operators may
be arbitrary functions of the remaining fields
$\{\phi,\bar\phi,\bar\rho,\psi,\bar\psi\}$ together with arbitrary
powers of their holomorphic derivatives (except that holomorphic
derivatives of $\bar\rho$ may be always be exchanged for other
fields using the $\rho$ equation of motion). The entire infinite
family of vertex operators is certainly of great interest,
interpreted in~\cite{Witten:2005px} as providing a
sheaf of chiral algebras over the target space $X$, while the
operators of conformal weight $(h,\bar h)=(0,0)$ form an interesting
generalization of the chiral ring of (2,2)
theories~\cite{Katz:2004nn,Adams:2005tc}.

Not all of these vertex operators will survive when we extend the
sigma model to a string theory in section~\ref{stringsec}. For
string theory, the key vertex operators are those which generate
deformations of the (0,2) moduli. These deformations are in
one-to-one correspondence with $\ov Q$-closed operators
$\cO^{(1,0)}$ of conformal weight $(h,\bar h)=(1,0)$ and charge +1
under $U(1)_R$, since given such an operator we can construct an
descendant $\int_\Sigma\cO^{(1,1)}\equiv\int_\Sigma\{\ov
Q^\dagger,\cO^{(1,0)}\}$ which satisfies
\begin{equation}
\left[\ov Q,\int_\Sigma\{\ov Q^\dagger,\cO^{(1,0)}\}\right]
=\int_\Sigma\left[\{\ov Q,\ov Q^\dagger\},\cO^{(1,0)}\right]
=\int_\Sigma\delbar\cO^{(1,0)}=0\ ,
\label{descent}
\end{equation}
because $\delbar=\rd$ when acting on sections of the canonical
bundle. Thus by its construction $\int_{\Sigma}\cO^{(1,1)}$ is
invariant under (0,2) supersymmetry, and if $\cO^{(1,0)}$ has
$U(1)_R$ charge +1 then $\cO^{(1,1)}$ will be uncharged, so that it
provides a marginal deformation of the worldsheet action\footnote{In
the A or B models the descent procedure may be taken one stage
further, relating deformations of the action to scalar operators of
vanishing conformal weight. But in (0,2) models there is only an
antiholomorphic supersymmetry so the descent procedure only affects
the antiholomorphic weight, mapping sections of $K^p\otimes \ov K^q$
to sections of $K^p\otimes\ov K^{q+1}$.}. As usual, these marginal
deformations are best interpreted as tangent vectors on the moduli
space of (0,2) models (at the base-point defined by the model in
question). We will have more to say on the role of finite
deformations in the twistor context in section~\ref{googlysec}.

Because $\bar\psi$ is a worldsheet (1,0)-form, operators of weight
$(h,\bar h)=(1,0)$ must be linear in either $\bar\psi_z$,
$\del_z\phi$, $\del_z\bar\phi$ or $\del_z\psi$. These fields are all
uncharged under $U(1)_R$, so if we want $\cO^{(1,0)}$ to have charge
+1 it must also be linear in $\bar\rho$. Then the only such
operators are
\begin{equation}
\begin{aligned}
&g_{i\bar k}\,\delta J(\phi,\bar\phi,\psi)^i_{\,\bar\jmath}\,\bar\rho^{\bar\jmath}\del_z\phi^{\bar k}&\qquad
&\bar\psi_{a\,z}\,\delta \j (\phi,\bar\phi,\psi)^a_{\,\bar\jmath}\,\bar\rho^{\bar\jmath}
\\
&\delta b(\phi,\bar\phi,\psi)_{i\bar\jmath}\,\bar\rho^{\bar\jmath}\del_z\phi^i&\qquad
&\delta\beta(\phi,\bar\phi,\psi)_{a\bar\jmath}\,\bar\rho^{\bar\jmath}\del_z\psi^a \ .
\end{aligned}
\label{vops}
\end{equation}
Note that $\delta J$, $\delta \j $, $\delta b$ and  $\delta\beta$
may depend arbitrarily on $\psi$ since it has $(h,\bar h)=(0,0)$,
although since $\psi$ is fermionic such dependence will be
polynomial. On the other hand, they must be independent of
$\bar\psi$ since this is a section of $K_\Sigma$. Each vertex
operator thus has a Taylor expansion in powers of $\psi$ and the
$p^{\rm th}$ coefficient of this expansion represents a section of
$\bigwedge^p \cV^\vee$. In particular, we can interpret the $U(1)_F$
quantum number as giving the transformation properties of the fields
under automorphisms of the line bundle $(\det\cV)^{1/{\rm rk}\cV}$,
whereupon the coefficients of the $\psi$ expansion  have $U(1)_F$
charge while the vertex operators as a whole are uncharged.
Geometrically, the fact that the $\psi$s are included in the vertex
operators in this way corresponds to the fact that the external
states should be thought of as wavefunctions on the supermanifold
$\bP^{3|4}$ that are holomorphic in the $\psi$s and may be expanded
as
\begin{equation}
f = \sum_{p=0}^4\,f_{i_1\cdots i_p}\psi^{i_1}\cdots\psi^{i_p}
\label{superfunction}
\end{equation}
where $f_{i_1\cdots i_k}$ is a section of $\bigwedge^p
\cO_{\bP^3}(-1)^{\oplus4}$. More abstractly, our presentation of
$\bP^{3|4}$ is as the space $\bP^3$ together with the structure
sheaf of superalgebras
\begin{equation}
\cO_{\bP^{3|4}}=\cO\left(\bigoplus_{p=0}^4\bigwedge\!\!^p\,\cO_{\bP^3}(-1)^{\oplus4}\right)\ ,
\label{superO}
\end{equation}
as in the standard abstract definition of a supermanifold (see {\it
e.g.}~\cite{Manin:1988ds,Deligne:1999qp}). The quantities $\delta J$
and $\delta  \j $ in the vertex operators~\ref{vops} can,
according to this interpretation, be indentified with a perturbation
of the almost complex structure of the supermanifold $\bP^{3|4}$
while $\delta b$ and $\delta\beta$ describe perturbations of the
$B$-field and hermitian structure on $\bP^{3|4}$.

The transformations~\ref{trans1} \& \ref{buntrans} show that $\ov Q$
acts on~\ref{vops} as
\begin{equation}
\ov Q = \bar\rho^{\bar\jmath}{\delta\ \over\delta\phi^{\bar\jmath}}\ ,
\label{dbarQ}
\end{equation}
in other words $\ov Q$ acts as the $\delbar$-operator on ${\rm
Maps}(\Sigma,X)$ (and restricts to the $\delbar$-operator on
instanton moduli space). Therefore, if \ref{vops} are to be
non-trivial in $\ov Q$-cohomology, $\delta J$, $\delta \j $,
$\delta b$ and $\delta \beta$ must represent (pullbacks to $\Sigma$
of) non-trivial elements
\begin{equation}
\begin{aligned}
&[\delta J]\in\bigoplus_{p=0}^4 H^{0,1}(X,T_X\otimes\bigwedge\!\!^p\,\cV^\vee)
&\qquad
&[\delta \j ]\in\bigoplus_{p=0}^4 H^{0,1}(X,\cV\otimes\bigwedge\!\!^p\,\cV^\vee)\\
&[\delta b]\in\bigoplus_{p=0}^4 H^{0,1}(X,T_X^\vee\otimes\bigwedge\!\!^p\,\cV^\vee)
&\qquad
&[\delta\beta]\in\bigoplus_{p=0}^4 H^{0,1}(X,\cV^\vee\otimes\bigwedge\!\!^p\,\cV^\vee)\ .
\end{aligned}
\label{vcoh}
\end{equation}
In fact, the interpretation of $\delta b$ is slightly more subtle.
$\delta b$ is defined upto the equivalence relation
\begin{equation}
\delta b \sim \delta b + \delbar\Lambda +\del M
\label{Bequiv}
\end{equation}
where $\Lambda\in \Omega^{1,0}(X)$ and $M \in\Omega^{0,1}(X)$. While
the freedom to add $\delbar\Lambda$ is the usual freedom in choice of
representative for a Dolbeault cohomology class, here we are also free
to add $\del M$ since $\del_i M_{\bar\jmath}\,\bar\rho^{\bar\jmath}
\del_z\phi^i=\del_z (M_{\bar\jmath}\,\bar\rho^{\bar\jmath})$ using the
$\bar\rho$ equations of motion, and so this term is a total
derivative. This corresponds to the fact that only the cohomology
class of $\cH=\del b\in H^1_{\delbar} (X,\Omega^2_{\rm cl})$
contributes to the moduli of a twisted (0,2) model.

If we take $X=\bP^3$, then because the Dolbeault complex is elliptic
and $\bP^3$ is compact, the above cohomology groups are at most
finite dimensional. Such cohomology corresponds via the Penrose
transform to fields on spacetime that extend over $S^4$ in the
Euclidean context (and indeed over the full compactified
complexification of Minkowski space, ${\rm Gr}_2(\bC^4)$). To obtain
fields on some subset of spacetime, we should take the target space
to be the noncompact region in twistor space swept out by the
corresponding lines. In the context of scattering theory, momentum
eigenstates extend holomorphically over affine complexified
Minkowski space $\bC^4\subset{\rm Gr}_2(\bC^4)$, the complement of
the lightcone at infinity. A suitable corresponding choice of target
subspace of twistor space is then $\bT'\equiv \bP^3-\bP^1$, and
$\bT'$ is isomorphic to the total space of the normal bundle
$\cO(1)+\cO(1)\to\bP^1$ of a line in $\bP^3$. More generally, one
could simply choose a tubular neighbourhood $\hat U$ of some fixed
line $L_p\subset\bP^3$, corresponding to a region $U$ around a
chosen spacetime point $p$. A particularly natural, conformally
invariant case is when $U$ is the {\it future tube}: the points of
complexified Lorentzian Minkowksi space for which the imaginary part
is timelike and future pointing, as this is the maximal domain of
extension of positive frequency functions. In this case, $\hat U$ is
the region $\bT^+$ on which the natural $SU(2,2)$-invariant inner
product is positive.

It is easy to see that the theory with noncompact target will remain
anomaly-free: we can naturally restrict the determinant line bundle
$\cL\to{\rm Maps}(\Sigma,\bP^3)$ to a line bundle over ${\rm
  Maps}(\Sigma,\bT')$, say, and the restricted bundle will be flat
since $\cL$ itself is. With this target space understood, via the
Penrose transform $\delta J$ describes an anti-selfdual $\cN=4$
conformal supergravity multiplet with helicities $-2$ to 0 (and
containing, in effect, two fields of helicity $-2$), $\delta \j $
describes four gravitino multiplets containing helicities
$-\frac{3}{2}$ to $+\frac{1}{2}$, while $\delta b$ and $\delta\beta$
are the CPT conjugates of $\delta J$ and $\delta \j $. From the
supermanifold point of view, $\delta J$ and $\delta \j $ combine to
describe deformations of the complex structure of $\bP^{3|4}$, while
$\delta b$ and $\delta\beta$ together represent deformations of the
cohomology class of the Kahler structure and NS flux on the
supermanifold, as detailed in~\cite{Berkovits:2004jj}.

\section{Coupling to Yang-Mills}
\label{ymsec}

We can incorporate Yang-Mills fields into the model by introducing a
worldsheet current algebra. This could be represented by adding in
further left-moving fermionic fields as in standard heterotic
constructions, or by a gauged WZW model, fibred over twistor space
as in~\cite{Distler:2007av}. For definiteness we will consider here
the simplest case of left-moving fermions
\begin{equation}
\lambda^\alpha\in\Gamma(\Sigma,\sqrt{K}\otimes\phi^*E)\quad\quad\quad
\bar\lambda_\alpha\in\Gamma(\Sigma,\sqrt{K}\otimes\phi^*E^\vee)
\label{YMferm}
\end{equation}
together with their (auxiliary) (0,2) superpartners. Here $E$ is a
rank $r$ holomorphic vector bundle over $\bP^3$ and, in contrast to
the $\psi$ fields, we have taken the $\lambda$s to be spinors on
$\Sigma$. The (0,2) transformations and action of these fields take
exactly the same form as in equations \ref{buntrans}-\ref{act2},
although the connection $\ov D$ acts now on sections of
$\sqrt K\otimes\phi^*E$, rather than just $\phi^*E$.

There are restrictions on $E$ arising from the requirement that this
coupling to $E$ does not disturb the anomaly cancellation in
section~\ref{anomsec}. All components of the quantum stress-tensor
will remain $\ov Q$-closed provided that the curvature $F^{(E)}$ of
the background connection on $E$ satsifies the Hermitian-Yang-Mills
equations $g^{i\bar\jmath}F_{i\bar\jmath}^{(E)}=0$. It is possible
to find such a connection~\cite{UY} if $E$ is stable and
\begin{equation}
\int_X {\rm c}_1(E)\wedge\omega\wedge\omega=0\ ,
\label{DUY}
\end{equation}
which for $X\simeq\bP^3$ implies that ${\rm c}_1(E)=0$ as
$H^{1,1}(\bP^3)$ is one-dimensional. Thus correlators in the $\ov
Q$-cohomology will conformally invariant at the quantum level if
${\rm c}_1(E)=0$ and $E$ is stable. Vanishing first Chern class of
the gauge bundle is a familiar condition in heterotic string
compactification, but it also plays a role in the Penrose-Ward
transform. A point in spacetime corresponds to a $\bP^1$ in twistor
space, so any twistor bundle that is the pullback of a spacetime
bundle must be trivial on every holomorphic twistor line, and this
will generically be the case provided $c_1(E)=0$.

In addition, ${\rm c}_1(E)=0$ ensures that there is an anomaly-free
$U(1)_{F'}$ global symmetry under which $\lambda$ and $\bar\lambda$
have equal and opposite charges and all other (dynamical) fields are
uncharged. Since this $U(1)_{F'}$ is conserved at the quantum level,
all correlation functions will vanish unless they involve equal
numbers of $\lambda$ and $\bar\lambda$ insertions.

\subsection{NS branes and Yang-Mills instantons}

Integrating out the non-zero-modes of $\lambda$ and $\bar\lambda$
provides a factor of $\det'(\delbar_{K^{1/2}\otimes\phi^*E})$ which
affects the sigma model anomaly, modifying the Green-Schwarz
condition to
\begin{equation}
0={\rm ch}_2(T_{\bP^3})-{\rm ch}_2(\cO(1)^{\oplus4})-{\rm ch}_2(E)\ .
\label{GS2}
\end{equation}
Since ${\rm ch}_2(T_{\bP^3})={\rm ch}_2(\cO(1)^{\oplus4})$, we must
require that ${\rm ch}_2( E)$ is trivial in $H^4(\bP^3,\bZ)$. Given
that ${\rm c}_1(E)=0$ for $E$ to be pulled back from a bundle over
spacetime,  \ref{GS2} requires further that $E$ is the pullback of a
Yang-Mills bundle with zero instanton number. Whilst it is
interesting to see how this well-known limitation of twistor-string
theory arises (which was not transparent in the original models), it
would be disappointing if twistor-string theory were truly
restricted to studying perturbative aspects of gauge theories.
Fortunately, the heterotic approach furnishes us with a mechanism to
avoid this constraint. At the non-perturbative level, heterotic
strings contain Neveu-Schwarz branes: magnetic sources for the NS
$B$-field. In the physical, ten-dimensional model, $B$ has a six-form
magnetic dual potential and the NS brane worldvolume is six
dimensional. However, in our six dimensional twisted theory the
magnetic dual of the $B$-field is again a two-form, so the twisted
theory contains NS branes with two dimensional worldvolumes,
wrapping curves $C\subset\bP^3$ that are holomorphic if the NS brane
does not break supersymmetry. If $[C]\in H^4(\bP^3,\bZ)$ is the
Poincar\'e dual of $C$, then the presence of a NS brane gives a
further contribution to the Green-Schwarz
condition~\cite{Duff:1996rs} which in our case reads
\begin{equation}
{\rm ch}_2(E)=[C]\ ,
\label{GSNSbrane}
\end{equation}
so that including NS branes wrapping holomorphic curves corresponds to
studying twistor-string theory in an instanton background.

In fact, the relation between Yang-Mills instantons and curves in
$\bP^3$ has long been known, and indeed was one of the earliest
applications of algebraic geometry to theoretical
physics~\cite{Atiyah:1977pw,Hartshorne:1978vv}. For example, to
construct the simplest case of an $SU(2)$ $k$-instanton described by
the 't Hooft ans\"atz\footnote{Here, $\sigma_{\mu\nu}$ is the
$su(2)$-valued anti-selfdual two form defined by
$\sigma_{ij}\equiv\frac{\im}{4}[\sigma_i,\sigma_j]$,
$\sigma_{0i}\equiv-\frac{1}{2}\sigma_i$.}~\cite{Jackiw:1976fs}
\begin{equation}
A(x)=\im\,{\rd x^\mu}\sigma_{\mu\nu}\del^\nu\log\Phi\ ,\qquad
\Phi(x)=\sum_{i=0}^k \frac{\lambda_j}{(x-x_i)^2}\ ,
\label{tHooft}
\end{equation}
one wraps NS branes on the $k+1$ lines $L_i\subset\bP^3$
corresponding to the points $x_i$ (with $x_0$ the point `at
infinity'). More specifically, each summand\footnote{The summands
$\Phi_i(x)$ are Green's functions for the scalar Laplacian on
spacetime, and are examples of twistor `elementary states'.} in
$\Phi(x)$ is represented on twistor space by $\tilde\Phi_i\in
H^1(\bP^3-L_i,\cO(-2))$ via the inverse Penrose transform. Similar
considerations hold for generic $SU(2)$
instantons~\cite{Atiyah:1977pw,Hartshorne:1978vv}, although it is
less clear how to extend the approach to higher rank gauge groups.

\subsection{Yang-Mills vertex operators}
\label{YMvopsec}

For the remainder of this paper, we will concentrate on Yang-Mills
perturbations around the zero-instanton vacuum. In a gauge in which
the background connection on $E$ vanishes, the (0,2) transformations
of $\lambda$ simplify to become
\begin{equation}
\delta\lambda^\alpha = \epsilon_2r^\alpha\qquad
\delta\bar\lambda_\alpha = \epsilon_1\bar r_\alpha\qquad
\delta r^\alpha = \epsilon_1\delbar\lambda^\alpha\qquad
\delta\bar r_\alpha = \epsilon_2\delbar\bar\lambda_\alpha,
\label{flatbuntrans}
\end{equation}
so that the action is
\begin{equation}
\left\{\ov Q,\int_\Sigma\!|\rd^2z|\ \bar\lambda_\alpha r^\alpha\right\}=
\int_\Sigma\!|\rd^2z|\ \bar\lambda_\alpha\del_{\bar z}\lambda^\alpha+\bar r_\alpha r^\alpha\ .
\label{flatbunaction}
\end{equation}
Thus the level one current algebra is represented as usual by free
fermions with propagator $\delta^\alpha_{\ \beta}/2\pi\im(z_1-z_2)$
in local coordinates $z$ on $\Sigma$. It is this current algebra
which is the natural heterotic realization of the current algebra on the worldsheet of Berkovits' twistor-string, or the current algebra of the D1-D5 strings in Witten's B-model twistor-string.

The coupling to $E$ provides new vertex operators of conformal
weight $(h,\bar h)=(1,0)$ and $U(1)_R$ charge 1, given by
\begin{equation}
\cO^{(1,0)}_\cA=\cA(\phi,\bar\phi,\psi)_{\bar\jmath\ \beta}^{\ \alpha}\,
\bar\rho^{\bar\jmath}\bar\lambda_\alpha\lambda^\beta
\label{YMvop}
\end{equation}
where again we allow $\cA$ to depend on $\psi$ but not
$\bar\psi$. This operator is non-trivial in $\ov Q$-cohomology
provided $\cA$ represents a non-trivial element of
$\bigoplus_{p=0}^4\,H^{0,1}(\bT',{\rm
End}E\otimes\bigwedge^p\cV^\vee)$ and represents a deformation of
the complex structure of $E\to\bP^3$, together with the $\cN=4$
completion. The integrated vertex operator corresponding
to~\ref{YMvop} is
\begin{equation}
\begin{aligned}
\cO^{(1,1)}_\cA&={\rm tr}\,\bar\lambda\left(\cA_{\bar\jmath}
\delbar\phi^{\bar\jmath}+\del_i\cA_{\bar\jmath}\rho^i\bar\rho^{\bar\jmath}
+\frac{\delta\cA_{\bar\jmath}}{\delta\psi^a}
A_{i\ b}^{\ a}\psi^b\rho^i\bar\rho^{\bar\jmath}\right)\lambda\\
&={\rm tr}\, \bar\lambda
\left(\phi^*\cA+D_i\cA_{\bar\jmath}\rho^i\bar\rho^{\bar\jmath}\right)\lambda
\end{aligned}
\label{Aint}
\end{equation}
up to terms proportional to the auxiliary fields, and where the trace
is over the Yang-Mills indices. The third term in the first line
arises through the $\psi$ dependence of $\cA$ and involves the
background connection $A$ on $\phi^*\cV$. Because
$\cV=\cO(1)^{\oplus4}$ is a sum of line bundles, we can always
choose this connection to be diagonal $A_{i\ b}^{\
a}=A_i\,\delta^a_{\ b}$. The second line, with $D$ the holomorphic
exterior derivative on sections of $\oplus_{p=0}^4\phi^*\left({\rm
End}\,E\otimes\bigwedge^p\cV\right)$, then follows since $\cA$ can
depend only polynomially on the fermions $\psi$. As expected,
comparing \ref{Aint} to \ref{act2} shows that
$\int_\Sigma\cO_\cA^{(1,1)}$ provides an infinitesimal deformation
of the worldsheet action corresponding to an infinitesimal change in
background super Yang-Mills connection.

To summarize, we have found a twisted (0,2) sigma model whose path
integral localizes on holomorphic maps to twistor space. Under the
Penrose transform, the tangent space to the moduli space of such a
(0,2) model corresponds to states in $\cN=4$ conformal SUGRA and
SYM, linearized around a flat background. For the SYM states,
introducing NS branes allows us also to discuss linearized
perturbations around an instanton background. However, the model
really contains an infinite number of other vertex operators that we
have not discussed, and at present there is no fully satisfactory
descent procedure relating deformations of the action to scalar
vertex operators. We will see that these issues are resolved when we
promote the sigma model to a string theory in the next section.
Moreover, whilst we were free to include the an additional
left-moving current algebra to describe a SYM multiplet, nothing in
the formalism has yet forced us to make a specific choice.

\section{Promotion to a String Theory}
\label{stringsec}

The (0,2) sigma model of the previous section depends on the choice of a
complex structure on $\Sigma$. This entered right at the beginning
in the definition of the (0,2) supersymmetry
transformations~\ref{trans1} \& \ref{buntrans}.  A choice of complex
structure on $\Sigma$, together with $n$ marked points to attach
vertex operators, is a choice of a point in the moduli space
of stable\footnote{We allow the abstract worldsheet to have nodes.}
curves $\ov\cM_{g,n}$ and to promote the sigma model to a string
theory, we should integrate over this space also.

In a twisted (0,2) model, as in the A or B
models~\cite{Witten:1992fb,Bershadsky:1993cx}, right-moving
worldsheet supersymmetry allows us to construct a top
antiholomorphic form on $\ov\cM_{g,n}$. Specifically, at genus $\geq
2$ we choose $3g-3+n$ antiholomorphic Beltrami differentials
$\ov\mu^{(i)}\in \ov{H^{0,1}(\Sigma,T_\Sigma)}$ and construct a
fermionic operator via the natural pairing $(\ov\mu^{(i)},\ov
G)\equiv\int_\Sigma \ov\mu^{(i)}\lrcorner\ \ov G$ with the (0,2)
supercurrent $\ov G=g_{i\bar\jmath}\rho^i\delbar\phi^{\bar\jmath}
\in\Gamma(\Sigma,\ov K\otimes\ov K)$. Inserting the product of
$3g-3+n$ such operators into the correlation function then provides
a top antiholomorphic form on $\ov\cM_{g,n}$.

In a twisted (2,2) model, the same procedure may also be used to
construct a top holomorphic form from the left-movers, but in our
(0,2) model we have no holomorphic supercurrent. Instead, we
introduce a holomorphic $bc$ ghost system (with apologies for possible
confusion with the $b=b_{i\bar j}$ field introduced earlier), with
\begin{equation}
b\in\Gamma(\Sigma,K\otimes K)
\qquad\qquad
c\in\Gamma(\Sigma,T_\Sigma)
\label{bcdef}
\end{equation}
having the natural action $S_{bc}=\int_\Sigma b\delbar c$. We will
take both $b$ and $c$ to be annihilated by the (0,2) supercharges
$\ov Q$ and $\ov Q^\dagger$. As in the bosonic (or left-moving
sector of the heterotic) string, including holomorphic $bc$ ghosts
provides us with a holomorphic BRST operator $Q$ such that the
holomorphic stress-energy tensor $T+T^{bc}$ of the sigma-model plus
$bc$ system is $Q$-exact, $T_{zz}+T^{bc}_{zz}=\{Q,b_{zz}\}$. In
parallel to the discussion above, a top holomorphic form on
$\ov\cM_{g,n}$ may be constructed from the $b$ antighosts by
inserting the product of $3g-3+n$ operators
$(\mu^{(i)},b)=\int_\Sigma \mu^{(i)}\lrcorner\ b$ into the path
integral. Of course, a proper treatment of a twisted (0,2) string
theory really requires an understanding of twisted versions of the
worldsheet (0,2) supergravity
of~\cite{Bergshoeff:1985gc,Brooks:1987nt}, just as the A and B model
topological strings may be understood from twisted (2,2)
supergravity~\cite{Verlinde:1990ku,Dijkgraaf:1990qw}.

\subsection{Constraints on the gauge group}
\label{c28sec}

The holomorphic BRST operator is nilpotent provided the left-moving
fields have vanishing net central charge. As in Berkovits'
model~\cite{Berkovits:2004hg}, this requires that the Yang-Mills
current algebra contributes $c=28$. This constraint arises from
integrating out the non-zero modes of $\{\phi,\rho,\psi,b,c\}$ and
the current algebra fields. If we represent the current algebra in
terms of left-moving fermions $\lambda$ as in section~\ref{ymsec},
we obtain a ratio of determinants\footnote{The presence of this
ratio is really a feature of (0,2) models; in a twisted (2,2) model
$\cV=T_X$ while there is no extra gauge bundle $E$ or $bc$ system,
so \ref{trivial} would automatically be unity. (0,2) supersymmetry
is sufficient to ensure that the ratio depends only holomorphically
on the moduli (as it ensures we only have determinants of
$\delbar$-operators), but the condition that \ref{trivial} be a
section of a flat line bundle becomes a non-trivial requirement.}
\begin{equation}
\frac{\det'\delbar_{\phi^*\cV}\,\det'\delbar_{\sqrt K\otimes\phi^*E}\,\det'\delbar_{T_\Sigma}}{\det'\delbar_{\phi^*T_X}}
\label{trivial}
\end{equation}
in the genus $g$ partition function. As in section~\ref{anomsec},
for $X=\bP^3$ and $\cV=\cO(1)^{\oplus4}$, the Quillen connection on
this determinant line bundle has curvature\footnote{The second line
in \ref{stringdet} follows if $E$ is trivial. In the presence of a
Yang-Mills instanton, the Quillen connection is not flat, but there
is a modification constructed from the NS field $H$ which
is~\cite{Freed:1986zx}.}~\cite{Bismut:1986ab,Freed:1986zx,Quillen}
\begin{equation}
\begin{aligned}
F&=\int_\Sigma\left.{\rm Td}(T_\Sigma){\rm ch}(T_\Sigma)
+{\rm Td}(T_\Sigma)\phi^*{\rm ch}(\cO(1)^{\oplus4}\ominus T_{\bP^3}))
+\hat A(T_\Sigma){\rm ch}(\phi^*E)\right|_{(4)}\\
&=\int_\Sigma\left.\left(1+\frac{x}{2}
+\frac{x^2}{12}\right)\left(2+x+{x^2\over 2}\right)
-\frac{x^2}{24}\,{\rm rk}\,E\,\right|_{(4)}\\
&=(28-{\rm rk}\,E)\int_\Sigma \frac{x^2}{24}
\end{aligned}
\label{stringdet}
\end{equation}
where $x={\rm c}_1(T_\Sigma)$. So for a current algebra at level one
we would require that $E$ has rank 28 as a complex vector bundle in
order to ensure that the determinant line bundle is flat and the
section \ref{trivial} may be taken as constant. More generally, a
current algebra at level $k$ contributes a central charge $c=k\,{\rm
rk}\,G/(k+h(G))$ for each semisimple factor $G$ of the Yang-Mills
gauge group, where $h(G)$ is the dual Coxeter number of $G$.

We have recovered the same constraint on the central charge of the
current algebra as in Berkovits' model~\cite{Berkovits:2004hg}. As
pointed out in~\cite{Berkovits:2004jj}, this is a rather puzzling
result. In conformal supergravity an $SU(4)$ subgroup of the $U(4)$
$R$-symmetry group is gauged\footnote{The remaining $U(1)$ factor is
the $U(1)_F$ symmetry acting on $\psi$ and $\bar\psi$,
responsible for the `helicity {\it vs} degree' selection rule
\ref{MHV2}.}. Spacetime field theory calculations by R\"omer \& van
Nieuwenhuizen~\cite{Romer:1985yg} show that this gauged $SU(4)_R$ is
anomalous unless the conformal supergravity is coupled to an $\cN=4$
SYM multiplet with gauge group either $U(1)^4$ or $U(2)$. We may
view this result as analogous to the statement~\cite{Green:1984sg}
that $\cN=1$ Poincar{\'e} supergravity in ten dimensions is
anomalous unless coupled to $\cN=1$ SYM with gauge group either
$U(1)^{496}$, $E_8\times U(1)^{248}$, $E_8\times E_8$ or
$Spin(32)/\bZ_2$. However, the small admissible gauge groups
$U(1)^4$ and $SU(2)\times U(1)$ in the conformal theory do not sit
well with the requirement that the Yang-Mills current algebra
contributes central charge 28, irrespective of the level $k$. In
contrast, for the physical heterotic string the required bundle
contribute central charge of 16 is perfectly tailored to the rank of
$E_8\times E_8$ or $Spin(32)/\bZ_2$.  Possible resolutions discussed
in~\cite{Berkovits:2004jj} include changing the level of the current
algebra or trying to include additional worldsheet fields without
changing the BRST cohomology.\footnote{It is perhaps worth noting
that, if it is possible to promote the sigma model to a string
theory without including a $bc$ system (as in the antiholomorphic
sector), then the net holomorphic central charge vanishes provided
the current algebra contributes $c=2$. This would be in better
accord with the required gauge groups. However, we do not know how
to do this.}

In the physical heterotic string, the requirement that the
determinant line bundle~\ref{stringdet} has trivial holonomy over
the moduli space of complex structures on $\Sigma$ fixes the gauge
group~\cite{Witten:1985mj,Bismut:1986wr}. (At genus 1, this amounts
to checking that the string partition function is invariant under
modular transformations of $\Sigma$.) We anticipate that modular
invariance will play a similarly important role in the context of
twistor-strings, and will likely rule out many solutions of
the central charge condition.

\subsection{Vertex operators in the string theory}
\label{stringvopssec}

When $Q^2=0$, there is a left-moving BRST complex graded by ghost
number, where $b$ and $c$ have ghost numbers $-1$ and $+1$,
respectively. As in section~\ref{vertexsec}, the relation
$\{Q,b_0\}=L_0$ shows that the $Q$-cohomology vanishes except for
states of holomorphic conformal weight $h=0$. Moreover, as in the
bosonic string, physical states are created by vertex operators of
ghost number $+1$. Since $c\in\Gamma(\Sigma,T_\Sigma)$, to construct a
(reparametrization invariant) vertex operator with $h=0$ we must
couple $c$ to a sigma-model vertex operator of conformal weight
$(h,\bar h)=(1,0)$. These are the operators of equations~\ref{vops}
\&~\ref{YMvop}. The fact that, when coupled to the $bc$ system, only
these vertex operators remain out of the entire sheaf of chiral
algebras is the real reason for having singled them out in the first
place.

The relation $\{Q, b_{-1}\}=L_{-1}$ now enables us to complete the
descent procedure: given an operator $\cO^{(p,q)}$ obeying
$\{Q,\cO^{(p,q)}\}=0$ we find that $\{b_{-1},\cO^{(p,q)}\}$ has
conformal weight $(p+1,q)$ and is $Q$-closed upto a total
holomorphic derivative. Consequently, there is now a complete
descent procedure between scalar vertex operators and deformations
of the worldsheet action.

\subsection{Contour integration on $\ov\cM_{g,n}(\bP^3,d)$}
\label{contoursec}

To compute scattering amplitudes involving $n$ external states, we
pick $n$ marked points on $\Sigma$ and attach a fixed vertex
operator for the appropriate external state to each. As usual, there
is an anomaly in the ghost number of the $bc$ system, given by the
excess of $c$ over $b$ zero-modes
\begin{equation}
h^0(\Sigma,T_\Sigma)-h^1(\Sigma,T_\Sigma)=3-3g\ .
\label{bcanom}
\end{equation}
This anomaly is completely absorbed by the $n$ vertex operators
and $3g-3+n$ factors of $(\mu^{(i)},b)$.

In the antiholomorphic sector however, the anomaly calculation \ref{Ranom}
showed that correlation functions vanish unless they contain
net $U(1)_R$ charge
\begin{equation}
h^0(\Sigma,\phi^*T_{\bP^3})-h^1(\Sigma,\phi^*T_{\bP^3}) = 4d + 3(1-g)\ .
\label{Ranom2}
\end{equation}
Since $\ov G_{\bar z\bar z}$ and the vertex operators have $U(1)_R$
charges $-1$ and $+1$ respectively, the insertion
$\prod^{3g-3+n}(\ov\mu^{(i)},\ov G)$ together with the $n$ vertex
operators contribute net $U(1)_R$ charge $3(1-g)$, but an anomaly of
$4d$ still remains\footnote{Note that this issue is not resolved
merely by moving to a model with $\bP^{3|4}$ target; one then finds
$h^0(\Sigma,\phi^*T_{\bP^{3|4}})-h^1(\Sigma,\phi^*T_{\bP^{3|4}})=-(1-g)$.}.
This residual anomaly - arising from an excess of $\bar\rho$
zero-modes - has a simple interpretation. Upon transforming the
fixed vertex operators to integrated ones using the
$(\ov\mu^{(i)},\ov G)$ insertions we are left with an integral over
the moduli space $\ov\cM_{g,0}(\bP^3,d)$ of degree $d$ stable maps
to $\bP^3$. This space has virtual dimension
\begin{equation}
{\rm vdim}\,\ov\cM_{g,0}(\bP^3,d)
=\int_{\beta}{\rm c}_1(T_{\bP^3})
\,+\,\left({\rm dim}_\bC \bP^3-3\right)\,(1-g)=4d\ .
\label{vdimKont}
\end{equation}
Consequently, the twistor-string path integral reduces to an
integral over a $4d$-dimensional moduli space (when the map is
unobstructed and $d>0$) in contrast to the case of a Calabi-Yau
target where the moduli space is (virtually) a discrete set of
points. This positive dimension is of course fully expected; in
particular $\ov\cM_{0,0}(\bP^3,1)={\rm Gr}_2(\bC^4)$, the conformal
compactification of complexified flat spacetime. Integrating out all
the fermion zero-modes, except the $4d$ `excess' $\bar\rho$ zero
modes, provides us not with a top form on $\ov\cM_{g,0}(\bP^3,d)$,
but instead a section of the canonical bundle\footnote{This section
is constructed from the $\psi$ zero-modes, representing a section of
the canonical bundle of instanton moduli space as in
section~\ref{globanomsec}, and the $bc$ zero-modes, furnishing a
section of the canonical bundle of the moduli space of curves.}
$\Omega^{4d,0}$. Such a form is most naturally integrated over a
real slice of $\ov\cM_{g,0}(\bP^3,d)$, which at $g=0$ and $d=1$ is
just a real slice of complexified spacetime. Indeed, on physical
grounds it is entirely appropriate that amplitudes should arise from
integrals over the real slice of spacetime rather than its
complexification.

A natural way to find a contour is to choose real structures, {\it
i.e.} antiholomorphic involutions $\tau_{\bP^3}:\bP^3\to\bP^3$ and
$\tau_\Sigma:\Sigma\to\Sigma$ obeying
$\tau_{\bP^3}^2=1=\tau_\Sigma^2$. These induce a real structure
$\tau$ on $\ov\cM_{g,0}(\bP^3,d)$ by
$\tau(\phi)=\tau_{\bP^3}\circ\phi\circ\tau_\Sigma$. The contour is
then the locus of maps invariant under $\tau$, so that
$\tau\phi=\phi$. This method was used by Berkovits
in~\cite{Berkovits:2004hg} to define twistor strings for spacetime
of signature $(++--)$, where $\tau_{\bP^3}$ and $\tau_\Sigma$ act by
standard complex conjugation on the homogeneous coordinates of the
target space and worldsheet. These choices of real structure leave
fixed an $\bR\bP^3$ submanifold of twistor space and an equatorial
$S^1\subset\Sigma$ at genus zero. In this case, real maps ({\it
i.e.} those left fixed by $\tau$) must take marked points of
$\Sigma$ to the fixed slice in twistor space so that vertex
operators are inserted on this fixed slice, as in Berkovits' model.
The same contour was used in the explicit calculations of Roiban,
Spradlin \& Volovich~\cite{Roiban:2004yf}

It would be desirable not to be reliant on split signature.
Calculations in split signature give satisfactory answers at tree
level, but it is thought that they will not straightforwardly extend
to loop amplitudes because the $\im\epsilon$ prescription for the
Feynman propagator will not be properly incorporated. Euclidean
spacetime signature corresponds to the real structure on $\bP^3$
given by quaternionic conjugation of the homogeneous coordinates. At
genus zero, one can combine this conjugation with the antipodal map
on the Riemann surface\footnote{The real structure also extends
beyond genus zero, as is most easily seen by considering the higher
genus Riemann surface as a branched cover over $\bP^1$, with pairs
of branch points at mutually antipodal points.} to give a real
structure on $\ov\cM_{g,n}(\bP^3,d)$. When $g=0$ and $d=2k+1$ this
method works well, but when $d=2k$ the fixed locus is empty.

For Lorentz signature, the reality conditions map twistor space to
dual twistor space and so do not define a real structure on $\bP^3$
in the same way as above, but instead give a pseudo-Hermitian
structure of signature $(2,2)$ on the non-projective twistor space.
The real points of Lorentz signature spacetime correspond to those
degree one rational curves in twistor space that lie in the zero-set
$\bN$ of the Hermitian form. However, connected curves of higher
degree are not likely to lie in $\bN$. Thus, in neither of these
physically more useful signatures are we able to obtain a
canonically defined real slice of the moduli space of stable maps.

One can avoid these problems if one is allowed to consider
disconnected curves, as, in the Euclidean case, a curve of even
degree can be represented as the union of two real curves of odd
degree, while in the Lorentzian case, one can simply make up a
degree $d$ curve as a union of $d$ degree 1 lines in $\bN$. Allowing
disconnected curves essentially entails moving to string field
theory, and this is discussed in section~\ref{sftsec}.

However, to make sense of twistor-string amplitudes in Euclidean and
Lorentzian signature, one does not need to go into string field
theory. The key point is that the contour only needs to be defined
as a homology cycle supported in an appropriate subset of the moduli
space. According to the philosophy given in~\cite{Gukov:2004ei}, it
is natural to think of the moduli space of instantons of fixed
degrees, but with different numbers of components as being joined
across spaces of nodal curves, and it is natural to allow the
contour to pass through these loci of singular curves. Although the
integrands have simple poles at such singular loci, the residues are
the same from both sides. Thus we can define the contour canonically
at degree $d$ as the appropriate $d$-fold product of real spacetime
in the space of $d$-component degree one curves. Then we deform this
contour into the space of connected, degree $d$ curves through nodal
curves. Although such a deformed contour will be non-canonical, it
is reasonable to hope that its homology class will be.

However the contour is chosen, we must implement it in the path
integral. To do so, suppose first of all that the contour has
Poincar{\'e} dual $\Gamma\in\Omega^{4d}(\ov\cM_{g,0}(\bP^3,d))$, and
let $\{t^A\}$ be a set of coordinates on a local patch of instanton
moduli space $\cM$, where $A=1,\ldots,h^0(\Sigma,\phi^*T_X)$. Then
for any stable holomorphic map $\phi$, we may expand a $\bar\rho$
zero-mode as
\begin{equation}
\bar\rho^{\bar\jmath}=\bar\rho^{\bar A}\frac{\del\phi^{\bar\jmath}}{\del \bar t^{\bar A}}
\end{equation}
so that $\{\bar\rho^{\bar A}\}$ correspond to a basis of (0,1)-forms
on $\cM$. Projecting $\Gamma$ onto its $(0,4d)$-form part (as usual
for contour integrals) we insert the operator
$\cO_\Gamma=\Gamma_{\bar A_1\cdots\bar A_{4d}}\bar\rho^{\bar
A_1}\cdots\bar\rho^{\bar A_{4d}}$ at degree $d$, so that we compute
\begin{equation}
\left\langle\cO_\Gamma
\prod_{i=1}^{3g-3+n}(\mu^{(i)},b)(\ov\mu^{(i)},\ov G)\
\prod_{j=1}^n\cO^{(0,0)}_j\right\rangle
\label{corr}
\end{equation}
where $\cO_j^{(0,0)}$ is a fixed vertex operator, formed from the
contraction of a $c$ ghost with one of the sigma model vertex
operators in~\ref{vops} or~\ref{YMvop} for external states in the
conformal supergravity or super Yang-Mills multiplets, respectively.
The $\cO_\Gamma$ insertion is to be thought of as part of the
definition of the degree $d$ heterotic path integral measure.

At $g=0$ there are no zero-modes of $b$, $\rho$ or $\bar\psi$, so as
usual the $bc$ and $\rho\bar\rho$ OPEs may be used to replace $n-3$
of the fixed vertex operators and all the
$(\mu^{(i)},b)\,(\ov\mu^{(i)},\ov G)$ insertions in~\ref{corr} by
$n-3$ integrated vertex operators $\int_\Sigma\cO^{(1,1)}$,
leaving us with
\begin{equation}
\left\langle\prod_{i=1}^3\cO^{(0,0)}_i
\prod_{j=4}^n\int_\Sigma\cO^{(1,1)}_j\right\rangle_{\!\Gamma}
\label{g0corr}
\end{equation}
where the subscript $\Gamma$ indicates the choice of contour.

Let us assume that the external states are all from the Yang-Mills
supermultiplet. We now integrate out the $\bar\lambda\lambda$
current algebra. There are no holomorphic sections of
$K^{1/2}\otimes\bC^r$ at genus zero, so we must take account of the
$\lambda\bar\lambda$ insertions when integrating out their
non-zero-modes. A standard approach is to introduce a coupling
$\int_\Sigma\,{\rm tr}\bar\lambda J\lambda$ to an arbitrary source
$J$, and then replace the $\bar\lambda\lambda$ factors in the vertex
operators by functional derivatives with respect to $J$. The path
integral over $\lambda$s then gives $\delta^n/\delta J^n\ \det
(\delbar_{\sqrt K\otimes\phi^*E}+J)$, evaluated at $J=0$. We have
\begin{equation}
\delta \det(\delbar_{\sqrt K\otimes\phi^*E}+J)
=\quad{\det(\delbar_{\sqrt K\otimes\phi^*E}+J)\over 2\pi\im}
\int_\Sigma {\rm tr}\, G'_J(u,u)\ \delta J(u)
\label{multtrace}
\end{equation}
where $u$ are homogeneous coordinates on the $\bP^1$ worldsheet and
$G'_J=G_J-G_0$ is the regulated Green's function for the $\delbar+J$
operator, with
\begin{equation}
G_J|_{J=0}={1\over2\pi\im}{\langle u_2\,\rd u_2\rangle\over\langle u_1\, u_2\rangle}
\label{Greens}
\end{equation}
where $\langle u\, v\rangle=\epsilon_{ab}u^av^b$ is the
$SL(2,\bC)$-invariant inner product. (Regularing by subtracting the
singular part $G_0(u,u)$ does not affect higher variations, which do
not require regularization.) This procedure gives multi-trace
contributions to the genus zero amplitudes, as in all the known
twistor-string theories: further variations can either act on $G'_J$
(leading to a single-trace contribution) or else act on the
determinant producing multi-trace terms.
In~\cite{Witten:2003nn,Berkovits:2004jj} these multi-trace terms
were attributed to conformal supergravity, formed from a number of
pure Yang-Mills interactions strung together with propagators
associated to fields in the conformal supergravity multiplet. From
the heterotic perspective also, such interactions are inevitable
since upon cutting the worldsheet between the fixed Yang-Mills
vertex operators, unitarity demands that all the states in the BRST
cohomology\footnote{Subject to the usual selection rules}, including
the conformal supergravity modes, appear in the cut. Note that,
after turning off the external current, both the single-trace and
multi-trace terms are accompanied by a factor of
$\det(\delbar_{\sqrt K\otimes\phi^*E})$. This factor combines with
the integral over the non-zero-modes of $\phi$, $\rho$, $\psi$ and
the $bc$ system to yield the ratio~\ref{trivial}, which as discussed
before may be taken as a constant due to anomaly cancellation.

Identifying the tree-level SYM amplitude with the leading-trace
term and integrating out the three $c$ zero-modes one obtains
\begin{equation}
\int\![\rd\phi\rd\psi\rd\bar\rho]_0\,\Gamma \e^{-S_{\rm inst}}
\ {\rm tr}\left\{
\cA_{\bar\jmath_1}\bar\rho^{\bar\jmath_1}\,
\cA_{\bar\jmath_2}\bar\rho^{\bar\jmath_2}\,
\cA_{\bar\jmath_3}\bar\rho^{\bar\jmath_3}\prod_{p=4}^n\int_{\Sigma}
\frac{\langle u_p\,\rd u_p\rangle}{\langle u_p\,u_{p+1}\rangle}
\cA_{\bar\jmath_p}\delbar\phi^{\bar\jmath_p}\right\}\ ,
\end{equation}
plus non-cyclic permutations, where $u_{n+1}\equiv u_4$ and the
trace is over the Yang-Mills indices. Finally, integrating out the
$3+4d$ $\bar\rho$ zero-modes from the vertex operators and the
contour insertion reduces this to the same integral that was the
starting point for the amplitude calculations
in~\cite{Witten:2003nn,Roiban:2004yf}. We have thus shown that the
leading-trace contribution to the amplitudes of heterotic
twistor-strings coincide with those of Witten's B-model.

\section{The geometry of supertwistor spaces and googly data}
\label{googlysec}

We have quantized on a region in a homogeneous twistor space
$\bP^3$, coupled in different ways to bundles $\cV=\cO(1)^{\oplus
4}$ and a trivial bundle $E$. The vertex operators correspond via
the descent procedure to perturbations of the action that correspond
to deformations of the geometric structures on this space.  In
particular, the operators in the first line of \ref{vops} were seen
to correspond to integrable deformations of the complex structure
$\cJ=(J,j)$ on the supermanifold $\bP^{3|4}$ and the second line to
$\delbar$-closed deformations of a NS field $\cB:=(b,\beta)$. Thus,
as reviewed in sections~\ref{vertexsec} \&~\ref{stringsec}, the
physical states of (heterotic) twistor-string theory are in
one-to-one correspondence with elements of the cohomology groups
$H^1(\bT'^{3|4},T_{\bT'^{3|4}})$, $H^1(\bT'^{3|4},\Omega^2_{\rm
cl})$ and $H^1(\bT'^{3|4},{\rm End}\,E)$. In turn, these groups
correspond via the Penrose transform to supermultiplets in $\cN=4$
conformal supergravity and super Yang-Mills, but it is important to
note that they represent only {\it linearized perturbations} around
some fixed background. For example, in the gravitational sector the
group $H^1(\bT'^{3|4},T_{\bT'^{3|4}})$ contains states describing
fluctuations of helicities $-2$ upto $+1/2$ that constitute the
anti-selfdual half of the spectrum of linearized $\cN=4$ conformal
supergravity. Going beyond perturbation theory, one first identifies
$H^1(\bT',T_{\bT'})$ as the tangent space to the moduli space of
complex structures on twistor space, and then Penrose's non-linear
graviton construction~\cite{Penrose:1976jq} states that a finite
deformation of the complex structure on $\bT'$ corresponds to a
four-dimensional spacetime with vanishing selfdual Weyl tensor
$W^+=0$.

The fact that perturbations of $\cJ$ and $\cB$ only have holomorphic
dependence on $\psi^a$ is not a restriction because a general
complex supermanifold $\cM_s$ can be expressed as the parity reverse
of a holomorphic vector bundle $\cV$ over the body $\cM$ but with
$\delbar$-operator deformed by terms that depend holomorphically on
the anticommuting fibre coordinates $\psi^a$ of $\cV$.  Thus we
require that the antiholomorphic tangent bundle of $\cM_s$ be
spanned by vectors of the form
\begin{equation}
\left\{
\frac{\del}{\del\phi^{\bar\imath}}+J_{\bar\imath}^{\ j}\frac{\del}{\del\phi^j}
+\j^a_{\bar\imath}\frac{\del}{\del\psi^a}\ ,\frac{\del}{\del\bar\psi^{\bar a}} \right\}
\end{equation}
where $\cJ=(J,\j)$ depends only on
$(\phi^i,\bar\phi^{\bar\jmath},\psi^a)$ with $\psi^a$ taken to be
anticommuting; we never need to have non-trivial functional
dependence on $\bar\psi^a$.  That this is no restriction on the
class of supermanifolds considered follows from the details of the
classification of complex supermanifolds in terms of cohomology on
the body~\cite{Eastwood:1986el,Manin:1988ds}. The above
representation corresponds to the situation in which the cohomology
classes are to be Dolbeault.

Similar considerations apply to the second line of \ref{vops}, which
corresponds to deformations of a supersymmetric extension $\cK=(K_i\rd
\phi^i\, , \kappa_a\rd\psi^a)$ of the form $K$ required to write the
action and its derivative
\begin{equation}
\cB=(b,\beta)= (K_{i,\bar\jmath}\, \rd\phi^i\wedge\rd\phi^{\bar\jmath},
\kappa_{a,\bar\jmath}\, \rd\psi^a\wedge\rd\phi^{\bar\jmath})\, .
\end{equation}
In the simplest case, $b$ and $\cB$ can be chosen to be global (note
that $K$ is not generally globally unless $\cH$ is
trivial).\footnote{The long exact sequence in cohomology that
follows from the short exact sheaf sequence
$$
0\to \cO/\bC\stackrel{\del}\to\Omega^{(1,0)}
\stackrel\del\to\Omega^{(2,0)}_{\mathrm{cl}}\to 0
$$
gives an obstruction in $H^2(\cO/\bC)$ for $\cH\in
H^1(\Omega^2_{\mathrm{cl}})$ to be written as $\cH=\del b$ for $b\in
H^1(\Omega^{(1,0)})$.  However, it can be seen that $H^2(\cO/\bC)=0$
in the twistor context: this follows from the long exact sequence in
cohomology arising from the sheaf sequence
$$
0\to \bC\to\cO\to\cO/\bC \to 0
$$
together with the vanishing of $H^3(\bC)$ and $H^2(\cO)$.  The first
of these vanishes because the twistor spaces for topologically
trivial spacetimes have topology $S^2\times \bR^4$ which has no
third cohomology.  The second follows for twistor spaces for Stein
regions in spacetime by the Penrose transform.}

One remarkable feature of twistor-string theory is that it gives a
partial resolution of the `googly problem'. As far as non-linear
constructions are concerned, this is the problem that while
anti-selfdual fields are understood fully nonlinearly in terms of
deformations of the complex structure of twistor space, it has not
been possible to understand fully nonlinear selfdual fields (one can
only incorporate them linearly).

Twistor-string theory only resolves the issue of the nonlinearities
associated to selfdual fields perturbatively, at least in a
holomorphic manner. In the case of Yang-Mills, the $\cN=4$
supersymmetry incorporates the selfdual part of the field into the
the same multiplet as the anti-selfdual part described by the
deformation $\cA$ of the $\delbar$-operator $\delbar_E$ on $E$.  In
the case of conformal supergravity, the anti-selfdual part of the
field and the selfdual part form two distinct super-multiplets, with
twistor data $\cJ$ and $\cB$.  These were shown to give rise
respectively to the anti-selfdual and selfdual parts of the standard
$\cN=4$ conformal supergravity multiplets in linear theory by
Berkovits and Witten~\cite{Berkovits:2004jj}.  The novel part as far
as twistor theory is concerned is in the encoding of the selfdual
part into $\cB$ which at the perturbative level, as discussed
earlier, should really be thought of as defining a class $\del\cB$
in $H^1(\bT'^{3|4},\Omega^2_{\rm cl})$. Thus the googly problem in
this context is to understand how to similarly exponentiate this
cohomology group. In the string theory, a vertex operator
representing a class in $H^1(\bT',\Omega^2_{\rm cl})$ has the
interpretation of deforming the target space by turning on flux of
the NS $B$-field. The appropriate framework for studying target
spaces with $B$-field flux, and thus twistor spaces of general
four-manifolds, would then appear to be the twisted generalized
geometry of Hitchin and
Gualtieri~\cite{Hitchin:2004ut,Gualtieri:2003dx}, in which
holomorphic objects $\{X+\xi\,,\,Y+\eta\}\in T\cM\oplus T\cM^\vee$
are closed with respect to the twisted Courant bracket
\begin{equation}
[X+\xi,Y+\eta]_{\rm TC}\equiv [X,Y] + \cL_X\eta-\cL_Y\xi
-{1\over2}\rd\left(\imath_X\eta-\imath_Y\xi\right) +\imath_X\imath_Y
H \label{Courant}
\end{equation}
rather than the Lie bracket. It is fascinating that generalized
geometry, of interest in compactifying physical string theory, also
appears to be an important ingredient in solving the googly problem
in twistor theory.

\section{Relation to other twistor-string models}
\label{relationsec}

We would now like to explain the relation of the heterotic
twistor-string constructed above to the twistor-string models of
Berkovits~\cite{Berkovits:2004hg} and Witten~\cite{Witten:2003nn}.

\subsection{The {\v C}ech-Dolbeault isomorphism and Berkovits' twistor-string}
\label{berkovitssec}

Berkovits' twistor-string has a first-order worldsheet action and is
usually viewed as a theory of open strings with boundary mapped to a
real slice of the target space. We will see that this
real slice arises through an orientifolding of a closed string
theory, appropriate only when the spacetime signature is $(++--)$,
rather than via D-branes. In fact, the relation of general twisted
(0,2) models to $\beta\gamma$-systems with a first-order action has
been explored already in~\cite{Witten:2005px} and we need
do little more here than apply these ideas to the case when the target
space is twistor space.

Consider a (0,2) model with its standard action
\begin{equation}
S=\int_\Sigma\!|\rd^2z|\  g_{i\bar\jmath}(\del_{\bar z}\phi^i\del_z\phi^{\bar\jmath}
+\rho^i_{\bar z}\nabla_z\bar\rho^{\bar\jmath}) 
+\bar\psi_{a\,z} D_{\bar z}\psi^a
+F_{i\bar\jmath\ b}^{\ a}\bar\psi_{a\,z}\psi^a\rho^i_{\bar z}
bar\rho^{\bar\jmath} \ ,
\label{02action}
\end{equation}
but where the target space is now taken to be a patch
$U\subset\bP^3$ that is homeomorphic to an open ball in $\bC^3$.
Because $U$ is contractible, the topological term
$\int_\Sigma\phi^*(\omega-\im B)$ necessarily vanishes. Also, $U$
admits a flat metric and since the $\ov Q$ cohomology is not
sensitive to the choice of metric, we are free to set
$g_{i\bar\jmath}=\delta_{i\bar\jmath}$. Likewise, since $\cV\to U$
is necessarily trivial, the background connection $A$ on $\cV$ may
also be chosen to be flat. Thus the (0,2) model over $U$ reduces to
the free theory
\begin{equation}
S=\int_\Sigma\!|\rd^2z|\ \delta_{i\bar\jmath}
(\del_{\bar z}\phi^i\del_z\phi^{\bar\jmath}
+\rho^i_{\bar z}\del_z\bar\rho^{\bar\jmath})
+\bar\psi_{a\,z}\del_{\bar z}\psi^a\ .
\label{02free}
\end{equation}

Non-trivial vertex operators correspond to elements of the Dolbeault
cohomology groups $H^{0,p}(U,\cS)$ where $\cS$ is the sheaf of
chiral algebras, but since $U$ is contractible these cohomology
groups vanish if $p>0$. Consequently, the only non-trivial vertex
operators are holomorphic sections of $\cS$ over $U$, represented in
the sigma model by operators which have the form\footnote{Recall
that the vertex operator must be independent of $\rho^i_{\bar z}$
and antiholomorphic derivatives of the fields since it must have
weight $\bar h=0$. Also, the $\rho$ equation of motion may always be
used to eliminate dependence on holomorphic derivatives of
$\bar\rho$.}
\begin{equation}
\cO(\phi^i,\del_z\phi^i,\del^2_z\phi^i,\ldots;\del_z
\phi^{\bar\jmath},\del^2_z\phi^{\bar\jmath},\ldots;
\psi^a,\del_z\psi^a,\del^2_z\psi^a,\ldots,\bar\psi_{a\,z},\del_z
\bar\psi_{a\,z},\ldots)\ .
\nonumber
\end{equation}
These vertex operators are independent of $\rho$ and $\bar\rho$, and
must depend holomorphically on $\phi$ so that they involve
$\phi^{\bar\jmath}$ only through its first and higher derivatives.
Therefore we may equally well obtain them from the
$\beta\gamma$-system
\begin{equation}
S_{\beta\gamma}
=\int_\Sigma\!|\rd^2z|\ \left(\beta_{i\,z}\del_{\bar z}\gamma^i
+ \bar\psi_{a\,z}\del_{\bar z}\psi^a\right)
\label{bgsystem}
\end{equation}
where $\gamma^i:=\phi^i$ and
$\beta_{i\,z}:=\delta_{i\bar\jmath}\del_z\phi^{\bar\jmath}$. Note
that the interpretation of $(\phi^i,\psi^a)$ as  holomorphic
coordinates on a supermanifold is once again manifest in this
$\beta\gamma$ picture.

To recover the higher cohomology groups $H^p(X,\cS)$ from this
$\beta\gamma$ system, we work with a quantum field theoretic
implementation of {\v C}ech cohomology. Let $\{U_\alpha\}$ be a
good\footnote{{\it I.e.} the covering $\{U_\alpha\}$ must be a Leray
cover of $X$, meaning roughly that nothing new arises on choosing a
finer subcover. See {\it e.g.}~\cite{BottTu,GH} for introductions to
{\v C}ech cohomology, \cite{HT,WW} for introductions in the context
of twisor theory and~\cite{Witten:2005px} for a discussion in the
context of (0,2) models and $\beta\gamma$ systems.} cover for $X$,
where $\alpha$ indexes the covering set. On each open set $U_\alpha$
we may construct a free $\beta\gamma$-system as in~\ref{bgsystem},
but to recover the sigma model globally we must ensure that these
free field theories glue together compatibly on overlaps
$U_\alpha\cap U_\beta$: as explained in {\it
e.g.}~\cite{Witten:2005px, Nekrasov:2005wg}, this entails that the
target space $X$ and bundle $\cV\to X$ obeys the same anomaly
conditions as found in section~\ref{anomsec}. If
$\cO_{\alpha_0\alpha_1\ldots\alpha_p}$ is a vertex operator which is
holomorphic in $\gamma$ when restricted to the $p$-fold overlap
$U_{\alpha_0}\cap U_{\alpha_1}\cap\cdots\cap U_{\alpha_p}$, the 
{\v C}ech cohomology group $H^p(X,\cS)$ is represented by a collection
of vertex opertators that obey the cocycle condition
$\rho_{\left[\alpha_0\right.}\cO_{\left.\alpha_1\alpha_2\ldots\alpha_{p+1}\right]}=0$ on $p+1$-fold overlaps, where $\rho_{\alpha}$ restricts a vertex
operator defined on $U_\beta$ to the intersection $U_\alpha\cap
U_\beta$, and the square brackets denote antisymmetrization. This
collection is defined modulo the equivalence relation
\begin{equation}
\cO_{\alpha_0\alpha_1\ldots\alpha_p}\,\sim\,
\cO_{\alpha_0\alpha_1\ldots\alpha_p}
+\sum_{k=0}^p (-1)^k\cO_{\alpha_0\ldots\widehat{\alpha_k}\ldots\alpha_p}
\label{cochain}
\end{equation}
for coboundaries, where
$\cO_{\alpha_0\ldots\widehat{\alpha_k}\ldots\alpha_p}$ is holomorphic
on the $(p-1)$-fold overlap $U_{\alpha_0}\cap\cdots\cap
U_{\alpha_{k-1}}\cap U_{\alpha_{k+1}}\cap\cdots\cap U_{\alpha_p}$ with
$U_{\alpha_k}$ omitted.

Rather than working with a covering of the projective twistor space,
we could equally well consider a `gauged $\beta\gamma$ system' of
maps $Z:\Sigma\to\bC^{4|4}$ with action
\begin{equation}
S= \int_\Sigma Y_I\ov D Z^I
\label{gaugedbgact}
\end{equation}
where $I=(\alpha,a)$ runs over the four bosonic and four fermionic
directions, while the kinetic operator $\ov D Z^I =(\delbar+A)Z^I$
gauges the $\bC^*$ symmetry so as to carry out the quotient
$\bP^{3|4}=(\bC^{4|4}-\{0\})/\bC^*$. It is straightforward to see
how these approaches are related: integrating out $A$ yields the
constraint $Y_IZ^I=0$ which may be solved on the patch $Z^0\neq0$ by
setting $Y_0=-(Y_iZ^i+Y_a Z^a)/Z^0$, where $i$ runs over the three
remaining bosonic directions. Substituting this
into~\ref{gaugedbgact} gives
\begin{equation}
\begin{aligned}
S_{\{Z^0\neq0\}}&=\int_\Sigma Y_i\delbar Z^i +Y_a\delbar Z^a
-\left(Y_iZ^i+Y_a Z^a\right)\left({\delbar Z^0\over Z^0}\right)\\
&=\int_\Sigma \beta_i\delbar\gamma^i +\bar\psi_a\delbar\psi^a
\end{aligned}
\end{equation}
where $\gamma^i=Z^i/Z^0$ and $\psi^a=Z^a/Z^0$ are affine coordinates
on the patch $Z^0\neq0$, whereas $\beta_i=Z^0Y_i$ and
$\bar\psi_a=Z^0Y_a$. For more general twistor spaces, the
non-projective twistor space is not flat and cannot be covered by a
single coordinate patch.

In order to promote either~\ref{bgsystem} or~\ref{gaugedbgact} to a
string theory, one must again introduce a holomorphic $bc$ system
and a worldsheet current algebra to ensure that the total central
charge vanishes. The associated BRST operator restricts the
interesting vertex operators to those formed from a $c$ ghost
contracted with a $\beta\gamma$ vertex operator of weight $h=0$,
just as in section~\ref{stringsec}. The path integral now involves
only the holomorphic coordinates $Z^I$ and is naturally treated as a
contour integral.

Berkovits' model~\cite{Berkovits:2004hg} is usually presented as a
theory of open strings with the boundary $\del\Sigma$ of the
worldsheet being mapped to the real slice $\bR\bP^{3|4}$ of
supertwistor space. His action is
\begin{equation}
S=\Re\left\{\int_\Sigma Y_I\ov DZ^I + b\del c\right\}
\label{Berkovitsact}
\end{equation}
together with a current algebra contributing central charge 28 to
both the left- and right-moving sectors. The fields obey the
boundary conditions
\begin{equation}
Z^I=\ov Z^{\bar I}\qquad Y_I=\ov Y_{\bar I}\qquad b=\bar b\qquad
c=\bar c
\label{boundarycond}
\end{equation}
on $\del\Sigma$. This action and boundary conditions can be turned
into a closed string theory by gluing $\Sigma$ to its complex
conjugate $\ov \Sigma$ along the boundary to form a compact Riemann
surface $\Sigma_D$: the `double' of $\Sigma$. By construction, we
have a complex conjugation $\Sigma_D\to \Sigma_D$ which interchanges
$\Sigma$ with $\ov\Sigma$ and fixes the boundary $\del\Sigma$. To go
in the reverse direction, start with an action
\begin{equation}
S=\frac12\int_{\Sigma_D} Y_I\ov D Z^I+b\delbar c + S_{\rm YM}
\label{Berkovitsact-closed}
\end{equation}
on the closed Riemann surface $\Sigma_D$, where $S_{\rm YM}$ here is
a holomorphic current algebra. Upon restricting the path integral to
maps for which $\ov{Z^I(\sigma)}=Z^I(\bar\sigma)$ ({\it i.e.} taking
an orientifold projection) and decomposing
$\Sigma_D=\Sigma\cup\ov\Sigma$, this action reduces to Berkovits'
model~\ref{Berkovitsact}-\ref{boundarycond}. Thus from our
perspective, viewing the Berkovits model as an open string is really
a way of `hardwiring' in a choice of contour. Starting from a closed
string picture enables one to choose other contours relevant for
other spacetime signatures, at least in principle. Nonetheless, it
is remarkable that the original Berkovits model automatically takes
care of this issue and provides a practical way of evaluating
scattering amplitudes on a real spacetime slice, even if this comes
at the cost of the flexibility one expects in a contour picture.

\subsection{Witten's twistor-string: D5-D5, D5-D1 and D1-D1 strings}
\label{wittensec}

Witten's model consists of an open string topological B-model 
coupled to D1-branes in supertwistor space $\bT_s$, a region in $\bP^{3|4}$. The D1-branes wrap holomorphic curves $C\subset\bT_s$, and the D1-D5 open strings are modelled by a pair of fermionic worldsheet spinors
\begin{equation}
\alpha\in\Gamma(C,S_-\otimes E)\qquad\qquad
\beta\in\Gamma(C,S_-\otimes E^\vee)
\end{equation}
with action $\int_C\beta\bar\del_E\alpha $ on the holomorphic curve $C$. Performing the $\alpha\beta$ path integral yields the determinant $\det\delbar_{E\otimes S_-}$ which depends on the complex structure of the bundle. In the original proposal~\cite{Witten:2003nn,Roiban:2004yf}, one seeks to obtain a generating functional for Yang-Mills scattering amplitudes by integrating this determinant over a contour in the moduli space of curves. Expanding $\det\delbar_{E\otimes S_-}$ in powers of a perturbation of the background connection on $E$ leads to multi-trace terms which were the first hint of a coupling to conformal supergravity~\cite{Witten:2003nn}. Welcome or not, if conformal supergravity is present one would expect to be able to describe scattering processes involving external conformal supergravity states, so it is clear that the above proposal cannot be the whole story.

What is lacking is a theory of the D1-D1 strings on the worldvolume of the D-instanton. This may be obtained by dimensional reduction from of the Abelian holomorphic Chern-Simons theory
\begin{equation}
S_{\rm D5}=\int_{\rm D5}\!\!\Omega^{3|4}\wedge\cA\wedge\delbar\,\cA
\label{d5act}
\end{equation}
on the worldvolume of a single D5-brane, as in~\cite{Aganagic:2000gs,Dijkgraaf:2002fc}. To dimensionally reduce this action, we take the D5 worldvolume to be the total space of the normal bundle $N_{C|\bT_s}$ to a fixed curve $C$, so that the tangent space to the D5 brane decomposes as $T_{\rm D5}= T_C\oplus N_{C|\bT}$. Similarly, the (0,1)-form $\cA$ decomposes as 
\begin{equation}
\cA\in\Gamma(D5,\ov T_C^\vee)\oplus\Gamma(D5,\ov N_{C|\bT_s}^\vee)
\end{equation}
and only the components in $\Gamma(D5,\ov N_{C|\bT_s}^\vee)$ survive in \ref{d5act} under the assumption that $\cA$ is constant along the normal bundle fibres. Integrating out these fibre directions then gives the action
\begin{equation}
S_{\rm D1}={\rm vol}(N)\int_C\! \Phi_1\,\delbar \Phi_0
\label{d1act}
\end{equation}
on the worldvolume of the D1-brane, where $\Phi_0\in\Gamma(C,N)$ and $\Phi_1\in\Gamma(C,K_C\otimes N^\vee)$.

Putting this together, integrating out the fluctuations of the D1-D1 and D1-D5 strings gives a net contribution
\begin{equation}
\frac{\det\delbar_{E\otimes S_-}(C)}{\det'\delbar_{N_{C|\bT_s}}(C)}
\label{d1sd5s}
\end{equation}
to the path integral for each curve $C$ that the D-instantons wrap. We now compare this to the ratio \ref{trivial} obtained by integrating out the non-zero modes of the heterotic string. Using the facts that $N_{\Sigma|\bP^3}=T_{\bP^3}/T_\Sigma$ and $T_{\bT_s}=T_{\bP^3}\oplus\Pi\cV$ shows that \ref{trivial} and \ref{d1sd5s} coincide, at least when the heterotic map $\phi:\Sigma\to\bP^3$ is an embedding. The full contribution of a degree $d$ map in the heterotic string also involves the string action evaluated on a worldsheet instanton, and is
\begin{equation}
\int_{\cM_d}\!\!\!\rd\mu\ \exp\left(-\frac{A(C)}{2\pi} +\im\int_C B\right)\,
\frac{\det\delbar_{E(-1)}}{\det'\delbar_{N_{C|\bT'_s}}}\ .
\label{superphys}
\end{equation}
where $A(C)$ is the area of the curve as measured by the
restriction of the K\"ahler form to $C$ (one may rewrite this
exponential in terms of $b=B+\im\omega$) and we have also integrated over the space of curves $\cM_d$ in supertwistor space using the natural measure $\rd\mu$ as described earlier. Expression of this type of familiar from `physical gauge' calculations of corrections to the spacetime superpotential in heterotic compactifications due to worldsheet instantons~\cite{Becker:1995kb,Witten:1999eg,Buchbinder:2002ic,Beasley:2005iu}. Thus, the B-model and heterotic calculations agree so long as the D1-branes in the B-model couple electrically to the $b$-field. Precisely this coupling was assumed in~\cite{Berkovits:2004jj} by an argument based on the Green-Schwarz mechanism, and has also arisen in the context of a conjectured S-duality in topological strings on a standard Calabi-Yau~\cite{Nekrasov:2004js}.

To summarize, we have seen that the D1-D1 strings of the B-model
describe perturbative deformations of holomorphic curves in
supertwistor space, and are thereby associated with (the
anti-selfdual) half of the conformal supergravity multiplet in
spacetime. The D1-branes themselves involve a coupling to the
$b$-field which provides the selfdual half. The entire D1/D5 system,
including all the strings stretched between them, is succinctly
captured by the heterotic model. It would be fascinating to
investigate this duality further in the context of standard
topological strings.

\section{String field theory and twistor actions}
\label{sftsec}

In this section we make contact with the twistor actions of
\cite{Mason:2005zm,Boels:2006ir,Boels:2007qn}. The basic idea is
that, with some reasonable assumptions, the complete string field
theory can be shown to be equivalent to certain actions on twistor
space, which can in turn be shown to reduce to versions of conformal
supergravity coupled to Yang-Mills on spacetime. Modulo the
assumptions that we have to make, this gives a proof of equivalence
between our heterotic twistor-string and a particular version of
$\cN=4$ conformal supergravity coupled to super Yang-Mills.

In order to simplify notation in this section we will work
with supermanifolds.  Thus supertwistor space $\bT_s$ will
in the flat case be a region in $\bP^{3|4}$.  For our purposes $\bT_s$
is the total space of $\cV$ with parity reversed fibres.  In the
context of string field theory, we must work off-shell which means
that, at least initially, we consider almost complex structures
$\cJ$ on supertwistor space $\bT_s$ that are not necessarily
integrable.  In the context of the earlier discussions of vertex
operators, $\cJ=(J, \j )$ and infinitesimal deformations of $\cJ$
correspond to the top family of vertex operators in (\ref{vops}).
Similarly, the lower family corresponds to a variation of the
complexified Kahler structure $\cB=(\omega+\im B,\beta)$ on $\bT_s$.  We
first seek to formulate the theory when the geometric background is
`off-shell'. That is, the almost complex structure $\cJ$ is not
necessarily taken to be integrable, while $\cB$ and $\cA$ are taken to
be arbitrary (so that the $\delbar$-operator on $E$ defined by
$\cA$ is not integrable).  We will, however, take the almost complex
structure $(J, \j )$ to define a Calabi-Yau almost complex structure
on the manifold $\bT_s$ in which the vector fields
$\del/\del\bar\psi^a$ are antiholomorphic.  The Calabi-Yau condition
in this non-integrable context is taken to mean that there is a
canonical isomorphism between $\Omega^{3,0}$ and $\left(\det
  \cV\right)^\vee$ and this defines a $(3|4,0|0)$ integral superform
$\Omega$.  In this almost complex situation, the form $\Omega$ cannot
be closed, but $\rd\Omega$ will have bosonic type $(2|4,2|0)\oplus
(3|3, 2|0)$ with no $(3|4,1|0)$ term.

We consider first the contribution of a single degree zero
instanton.  This reduces to an integral over constant maps to
supertwistor space and zero-modes of the worldsheet fields
$(c,\bar\rho)$.  In principal, one should construct the contribution
to the string field theory action by formulating the sigma model for
an off-shell $(\cJ,\cB,\cA)$ and integrating out the zero modes of
$c$ and $\rho$.  An easier route, as followed in
\cite{Berkovits:2004jj}, is to calculate the cubic amplitudes as
integrals of cubic expressions in $(\cJ,\cB,\cA)$ and their
derivatives, and then guess the quadratic terms required to make
these contributions geometrically natural.  This process led
Berkovits and Witten to the following top degree form on
supertwistor space $\bT'_s=\bC\bP^{3|4}-\bP^1$
\begin{equation}
\cL_0(\cJ,\cB,\cA) = \left(\phantom{\frac{1}{2}}\!\!\!\ 
{\rm CS}(\cA) + N(\cJ)\lrcorner\,\cB
+ {\rm CS}(\del\cJ)\right)\wedge\Omega\ ,
\label{deg0act}
\end{equation}
where ${\rm CS}(\cA)={\rm tr} (\frac12 \cA\wedge \rd\cA + \frac13
\cA^3)$ is the Chern-Simons 3-form constructed from $\cA$. $N(\cJ)$
is the Nijenhuis tensor of the almost complex structure $\cJ$ on the
supermanifold. It is a section of $T^{(1,0)}\otimes \Omega^{(0,2)}$
and may be thought of as $(\delbar)^2$. Then $N\lrcorner\,\cB$ is
the $(0,3)$-form obtained by contracting the vector field part of
$N$ into $\cB$ and skewing over the anti-holomorphic indices. Note
that $(N\lrcorner\,\cB)\wedge\Omega$ may also be represented as
$\cB\wedge\rd\Omega$. Finally, ${\rm CS}(\del\cJ)$ is the
Chern-Simons (0,3)-form associated to the $\delbar$-operator
naturally induced on the holomorphic tangent bundle of $\bT'_s$ by
$\cJ$.    The contribution of a single degree-zero instanton to the
string field theory action is then
$S_0[\cJ,\cB,\cA]=\int_{\bT_s}\!\cL_0(\cJ,\cB,\cA)$.
Although~\ref{deg0act} was originally arrived at via the Berkovits
and Witten string theories, we have seen that the formulae for
amplitudes is the same in our heterotic theory, so the procedure
will lead to the same expression for our theory also.

With a rescaling $b$ to fit in with earlier conventions, the
contribution of the degree 1 instantons is given in
equation~\ref{superphys} as
\begin{equation}
S_1[\cJ,\cB,\cA]=\int\!\rd^{4|8}x\
\exp\left(\int_C\! b\right)\, \nonumber
{\det\delbar_{E(-1)}\over\det'\delbar_{N_{C|\bT'^{3|4}}}}\ \ .
\label{superphys1}
\end{equation}
For worldsheet instantons of degree greater than or equal to one, as
discussed in earlier sections we are reduced to a half-dimensional
contour integral inside the moduli space of curves of degree $d$.
Gukov, Motl and Neitzke~\cite{Gukov:2004ei} have argued that the
contour can essentially be continued through the boundaries of the
moduli spaces of Riemann surfaces of different degrees of
connectedness, so long as propagators associated to the above degree
zero action are allowed between points on the different components
of the curve (these can be thought of as being associated to
degenerations of a degree $d$ curve with vanishingly thin necks
connecting points on the different components).  The contact terms
between different components are therefore taken care of by the
degree zero action and so the contribution of a degree-$d$ instanton
consisting of $d$ degree 1 components is simply the product of $d$
copies of the degree 1 contribution.

To see this we note that if $C=\cup_{i=1}^d \bP^1_{x_i}$, the
integrals over $C$ behave additively,
$\int_C=\sum_i\int_{\bP^1_{x_i}}$ and so the exponentials behave
multiplicatively; similarly the determinants behave
multiplicatively.  Since the $d$ copies of $\bP^1$ are
indistinguishable, the degree $d$ integral becomes
\begin{equation}
\frac1{d!} \int \prod_{i=1}^d \rd^{4|8}x_i \; \cL_1(x_i)
=\frac1{d!}\left(\int\!\rd^{4|8}x \; \cL_1(x)\right)^d
=\frac{(S_1)^d}{d!}\ .
\label{degreed}
\end{equation}
The total contribution must also be summed over the number $k$ of
degree zero components, as well as over $d$. These contributions
should be divided by the number $k!$ of permutations of the
indistinguishable degree zero components. Thus the overall
contribution of degree $d$ instantons can be written as
\begin{equation}
\sum_d\left\{\sum_{k}\frac{1}{k!}
\left(\int\!\cL_0\right)^k\frac{(S_1)^d}{d!}\right\}
=\exp\left(S_0+S_1\right)\ .
\end{equation}

In string field theory one considers disconnected string
worldsheets, so the above argument shows that it is natural to take
$S_0+S_1$ to be the string field theory action. These actions are
also actions on twistor space, with $S_0$ being local, but $S_1$
non-local.

Parts of the action $S_0+S_1$ have been studied elsewhere.  The
truncation to spin one and spin two fields was studied in
\cite{Mason:2005zm} and shown to provide twistor space actions that
give rise to standard Yang-Mills theory and conformal gravity on
spacetime (in that analysis, the determinant factors in
(\ref{superphys1}) were not incorporated. Presumably they do not
change the truncated theory). The fully supersymmetric case for
Yang-Mills theory was studied in~\cite{Boels:2006ir} (see
also~\cite{Nair:2005iv}) where it was shown that pure $\cN=4$ super
Yang-Mills theory corresponds to the twistor action
\begin{equation}
\int_{\bT'_s}\Omega\wedge{\rm CS(\cA)}
+\int\!\rd^{4|8}x\log\det\delbar_{E(-1)} \ .
\end{equation}
The non-local part of the action here involves
$\log\det\delbar_{E(-1)}$, rather than $\det\delbar_{E(-1)}$ which
would be the truncation of the above, but leads to multitrace terms
in the action. We do not know how to obtain such a term from string
theory. We have not yet followed through the full details of the
Penrose transform (along the lines
of\cite{Mason:2005zm,Boels:2006ir}) to find the spacetime action
that is equivalent to $S_0+S_1$ above and thereby check the
conjectures of Berkovits and Witten~\cite{Berkovits:2004jj}.

\section{Discussion}
\label{discusssec}

To date, twistor-string theory has mainly been used indirectly as a
source of inspiration for calculating gauge and gravitational
scattering amplitudes in
spacetime~\cite{Cachazo:2004kj,Brandhuber:2004yw,Britto:2004ap,Bern:2005iz}.
However, we find it difficult to believe that these structures in
gauge and gravity theories are simply coincidental, and would like
to argue that their existence gives strong new support to Penrose's
original twistor programme~\cite{Penrose:1999cw}. This programme
seeks to reformulate all of fundamental physics in terms of complex
analytic objects on twistor space, with the intention that twistor
space be in some way the primary arena for physics, in which quantum
gravity might make the most sense. The remarkable reformulation of
anti-selfdual gravitational~\cite{Penrose:1976jq} and
Yang-Mills~\cite{Ward:1977ta} fields in terms of deformations of the
complex structures of twistor space itself or of a bundle over
twistor space provided impressive early successes which motivated
this programme. As we discussed in section~\ref{googlysec}, these
twistor-string ideas have given new insight into the googly problem,
as well as providing a new avenue towards incorporating quantum
field theoretic ideas into the twistor programme.

Clearly, more work is required to discover what other twistor-string
theories can be constructed.  In particular, one would like to have
twistor-string theories that give rise to Poincar{\'e}
supergravities, or to pure super Yang-Mills, or that incorporate
other representations of the gauge and Lorentz groups.  Some steps
have been made in this direction~\cite{AbouZeid:2006wu,Bedford:2007qj}.  
It is clear from the calculations of section~\ref{stringsec} that enforcing modular invariance will play a key role in selecting the gauge group, and we would like to investigate this further. Finally, we saw that the heterotic string path integral is naturally treated as a contour integral. Such a contour integral interpretation is required to correctly derive the results of Roiban, Spradlin \& Volovich~\cite{Roiban:2004ka} for scattering amplitudes in the `connected prescription'. Witten has proposed that the equivalence between the connected and disconnected prescriptions might be understood in terms of a residue theorem~\cite{Beasley:2003fx} for a twisted heterotic string.
We hope that the work in this paper will provide further tools for studying these questions.

\section*{Acknowledgments}
The authors would like to thank Nathan Berkovits, Chris
Beasley, Jacques Distler, Lotte Hollands, Chris Hull, Edward Witten
and especially Rutger Boels for useful comments and discussion. LM is
partially supported by the EU through the FP6 Marie Curie RTN {\it
  ENIGMA} (contract number MRTN-CT-2004-5652) and through the ESF {\it
  MISGAM} network. DS is supported by a Mary Ewart Junior Research
Fellowship.  This work was stimulated by a reading of Witten's
paper~\cite{Witten:2005px}, and, with hindsight, a heterotic model is
implicit in remarks in Berkovits \& Witten~\cite{Berkovits:2004jj} and
in Witten's talk at the LMS Twistor-String Conference in Oxford,
January 2005 (http://www.maths.ox.ac.uk/~lmason/Tws).

\end{document}